\documentclass[pdflatex,sn-mathphys-num]{sn-jnl}

\usepackage{graphicx}%
\usepackage{multirow}%
\usepackage{amsmath,amssymb,amsfonts}%
\usepackage[title]{appendix}%
\usepackage{xcolor}%
\usepackage{textcomp}%
\usepackage{manyfoot}%
\usepackage{booktabs}%
\usepackage{algorithm}%
\usepackage{algorithmicx}%
\usepackage{algpseudocode}%
\usepackage{listings}%

\usepackage[section]{placeins}%
\usepackage{amsfonts}%
\usepackage{xfrac}%
\usepackage{fancyvrb}%

\newcommand{\p}{\partial}
\newcommand{\sig}{\sigma}
\newcommand{\al}{\alpha}
\newcommand{\ga}{\gamma}
\newcommand{\la}{\lambda}
\newcommand{\bol}{\mathbf}
\newcommand{\volume}{\mathbb{V}}
\newcommand{\ovl}{\overline}
\newcommand{\bsym}{\boldsymbol}
\newcommand{\dv}[2]{\frac{d#1}{d#2}}
\newcommand{\ddv}[2]{\frac{d^2#1}{d#2^2}}
\newcommand{\hdiff}[2]{\left.\dv{#1}{\xi}\right|_#2}

\begin{document}

\title[Article Title]{Linear Stability Analysis of Two-phase, Two-Component Flow in Porous Media}




\author*[1]{\fnm{Paulo Lee Kung} \sur{Caetano Chang}}\email{Paulo.Chang@student.uib.no}

\author[1]{\fnm{Kundan} \sur{Kumar}}\email{kundan.kumar@uib.no}

\affil*[1]{\orgdiv{Department of Mathematics}, \orgname{University of Bergen}, \orgaddress{\street{Allégaten 41}, \city{Bergen}, \postcode{5007}, \state{Vestland}, \country{Norway}}}

\abstract{Viscous fingering instabilities during fluid displacement in porous media can compromise the efficiency of applications such as enhanced oil recovery, CO$_2$ sequestration, and groundwater remediation. While extensive research exists on linear stability analysis for fully immiscible and fully miscible displacements, the intermediate case of partially miscible flow with limited mass transfer between phases remains largely unexplored. This study extends linear stability analysis to a two-phase, two-component system that accounts for gravity effects, fractional flow, capillary forces, mechanical dispersion, and interphase mass transfer, focusing on the case where a partially miscible gaseous fluid displaces a liquid. We formulate an eigenvalue problem to characterize instability growth rates and cutoff wavenumbers. The resulting ordinary differential equations have discontinuous coefficients at the transition from two-phase to pure-liquid flow, resulting in discontinuous eigenfunction derivatives. We derive jump conditions for the derivatives at this transition, and solve the eigenvalue problem using the matched initial value problem method. Results demonstrate that mass transfer has a predominantly stabilizing effect by reducing viscosity contrast and altering shock properties at the displacement front. This stabilizing influence is particularly pronounced for high viscosity contrasts and dampens gravity-induced instability in upward displacements. Mass transfer most significantly affects the perturbation growth rate, while its effect on the cutoff wavenumber is less pronounced. We identify a critical value for the dimensionless longitudinal dispersion coefficient where both growth rate and cutoff wavenumber are maximized, suggesting complex interactions between capillary forces and mechanical dispersion.}







\maketitle
\section{Introduction}\label{sec:intro}
In this paper, we are interested in addressing the problem of viscous fingering in two-phase, compositional displacement process, where the invading and displaced phases are not fully miscible, but mass transfer occurs between the phases (partially miscible flow). Our work focuses on the prediction of unstable modes of perturbation using linear stability analysis.

Flow displacement processes in porous media are critical to numerous industrial and environmental applications. In enhanced oil recovery operations, the injection of water, gas, or chemical agents displaces oil from reservoir rocks, directly impacting production efficiency and economic viability. Similarly, in carbon sequestration and storage (CCS) projects, CO$_2$ is injected into deep saline aquifers or depleted hydrocarbon reservoirs, where the stability of the displacement front determines plume migration patterns. Groundwater remediation efforts also rely on displacement processes to remove or contain contaminants in subsurface formations. In all these applications, the stability of the displacement front is essential for process efficiency, as unstable flows lead to \textit{viscous fingering}---a phenomenon where the displacing fluid penetrates the displaced fluid in preferential pathways, creating finger-like intrusions. This instability significantly reduces sweep efficiency, causes premature breakthrough of the displacing fluid, and results in substantial bypassing of the target fluid, diminishing the effectiveness of enhanced oil recovery, compromising CO$_2$ storage security, and limiting remediation success.

Unlike flow channelling, which arises from spatial heterogeneity in the porous medium, viscous fingering stems from variations in fluid properties---such as viscosity and density---across the displacement front. These property differences may result from displacement between two immiscible phases or from mixing of two fully miscible fluids \citep{homsy1987}. When a small, localized perturbation at the displacing front (with the displacing fluid advancing across the front) leads to steeper (negative) pressure gradient, local flow rate is increased, amplifying the initial perturbation and leading to instability. If, on the other hand, the perturbation reduces the local pressure gradient, the displaced fluid resists the local advance, and the displacing front is stabilized. The primary drivers of instability are, for most scenarios, adverse mobility and density ratios. Conversely, favourable mobility ratios or density contrasts act as stabilizing forces.

Besides viscous and gravity forces, dissipative forces also influence the growth rate and length scale of finger-like perturbations. In macroscopic, immiscible flow, the primary dissipative mechanism is capillary pressure, while in miscible flow, molecular diffusion and mechanical dispersion play this role. These dissipative forces counteract any gradients in composition (whether saturation or molecular concentration), and act as stabilizing mechanisms by ($i$) smoothing the compositional gradient across the displacing front, limiting the growth rate of perturbations, and ($ii$) by smoothing transverse gradients along the front, effectively dispersing small-scale perturbations---which, in the context of modal analysis, results in a wavelength cutoff value below which flow is stabilized.

Linear stability analysis provides a powerful mathematical framework for predicting the onset of viscous fingering and characterizing the early stages of instability development. The analysis for displacement processes typically begins with a base state---the one-dimensional displacement profile at the displacement front---and superimposes small-amplitude, wave-like disturbances in the transverse direction (normal mode analysis). By linearizing the governing equations around this base state, one obtains an eigenvalue problem whose solution yields the growth rate as a function of perturbation wavelength. A positive growth rate indicates a mode is unstable, with its amplitude growing exponentially over time, while negative growth rates indicate stable modes that decay over time.

Since the pioneering study by~\citet{hill1952}, extensive research has been published on viscous fingering. A classical review on the subject was given by~\citet{homsy1987}. More recently, \citet{pinilla2021} provided an extensive review on experimental and computational studies on miscible and immiscible viscous fingering. A review focusing on variational methods for numerical simulation of unstable miscible displacement was given by \citet{scovazzi12017}, and a shorter, but well organized review on immiscible viscous fingering was provided by \citet{salmo2022}. 

Linear stability analysis alone has generated substantial literature. Following the foundational work of \citet{saffman1958} and \citet{chuoke1959}, several works have addressed linear stability analysis for immiscible viscous fingering (see e.g., for a recent review \citep{salmo2022}), including notable contributions by \citet{hagoort1974}, \citet{yortsos1986}, \citet{yortsos1989}, \citet{chikhliwala1988}, \citet{riaz2004}, and \citet{daripa2008}. Similarly, extensive research exists for miscible displacement, with important examples including \citet{tan1986}, \citet{hickernell1986}, \citet{manickam1995}, \citet{dewit1997}, \citet{riaz2003}, and \citet{jangir2021}. An important subset of miscible displacement is gravity-driven instability---typically the result of concentration gradients producing adverse density gradients in the aqueous phase---with notable work by \citet{riaz2006}, \citet{vanduijn2019}, \citet{lasser2021}, and \citet{bringedal2025}.

The literature on viscous fingering in multiphase compositional displacement (i.e., partially miscible flow), however, is considerably more limited, particularly regarding linear stability analysis. Many practical applications involve displacement processes at conditions that are neither immiscible nor fully miscible, with limited mass transfer occurring between invading and displaced fluids such as CO$_2$-enhanced oil recovery below the miscibility pressure or carbon sequestration in saline aquifers. This underscores the importance of studying the viscous fingering in the context of multiphase compositional model. 

Most studies of unstable multiphase compositional flow address fingering through numerical simulation of the nonlinear governing equations, analysing instability either by visual inspection of compositional fields or through indirect metrics such as sweep efficiency. In this category, \citet{moortgat2016} examined the impact of several factors---including capillarity, molecular diffusion, flow rates, domain geometry and medium properties---on viscous fingering, and concluded that displacement in partially miscible conditions differs profoundly from the fully miscible case. In the context of linear stability analysis, the study by \citet{bringedal2025}, which examines evaporation-driven density instabilities in porous media, considers a model incorporating both immiscible displacement between two phases (air and water) and miscible transport of a solute (NaCl) within the aqueous phase. A key distinction from the present work, however, is that interphase mass transfer does not occur in their analysed case. We are aware of only one work, by \citet{tran2017}, that accounts for mass transfer between phases. The authors focused on CO$_2$ displacement of heavy oil under conditions that are predominantly immiscible, but with limited interphase mass transfer. Their analysis assumed that: ($i$) displacement occurs in an idealized step-saturation profile with constant volume fractions on either side of the displacement front; ($ii$) capillary forces are neglected; and ($iii$) mass transfer follows a prescribed function rather than being determined from phase-equilibrium relationships. Their results show that mass transfer of CO$_2$ into the oil phase has a stabilizing effect on the displacement by reducing both the front velocity and the viscosity gradient across the front---findings that will be shown to be in agreement with our own analysis. 

In the present work, we focus the analysis of linear stability to partially miscible displacement in a two-phase, two-component system with mass transfer between phases. In contrast to~\cite{tran2017}, our model accounts for: gravity forces; fractional flow and capillary forces in the two-phase region; variable viscosity and mechanical dispersion in single-phase regions. We limit our study to the common scenario where the invading fluid is gaseous and the displaced fluid is liquid. While the two-component system does not fully capture the physics of many compositional flow problems---especially when the liquid phase contains multiple volatile components---it provides a useful framework that considerably simplifies the mathematical description of the problem (see Section~\ref{sec:phase_eq}). By accounting for partial miscibility, we attempt to the bridge the gap between the well-studied limiting cases of immiscible and miscible displacement. To the best of our knowledge, this is the first work dealing a linear stability analysis of a genuine two-phase, two-component system taking into account interphase mass transfer.

For the idealized system where pressure variation is small compared to the system pressure and liquid density is weakly dependent on composition, our two-phase, two-component model becomes quite similar to the immiscible case solved by \citet{riaz2004}. The key differences introduced by mass-transfer effects in the two-component case are: ($i$) shock calculations deviate from the classical Buckley-Leverett approach; ($ii$) a single-phase flow regime appears downstream of the shock region; and ($iii$) flow functions become discontinuous at the transition from two-phase to single-phase flow. The basic steps of the linear stability analysis for the proposed model are:
\begin{enumerate}
    \item From the general equations for multiphase compositional flow, assuming thermodynamic equilibrium, derive the governing equations for the simpler case of a two-phase, two-component system with additional simplifying assumptions (see Section~\ref{sec:phase_eq});
    \item From the one-dimensional, purely advective case, obtain the shock front configuration following the procedure developed by \citet{orr2007};
    \item Determine the steady-state composition profile around the shock (the travelling-wave solution of the one-dimensional problem)---the \textit{base state};
    \item Derive the perturbation equations by decomposing the primary variables into base-state solutions and normal mode perturbations, then linearizing the flow functions around the base state---the \textit{eigenvalue problem};
    \item Solve the eigenvalue problem using the matched initial value problem (MIVP) method (adapted from \citep{chikhliwala1988}), which consists of:
        \begin{enumerate} 
            \item Calculate the far-field solutions of the linearized flow equations for a given wavenumber to obtain the initial conditions on the far-upstream and far-downstream sides, characterized by arbitrary constants;
            \item Integrate the linearized equations from both the far-upstream and far-downstream sides toward the matching point (the flow regime transition point) using the calculated initial conditions;
            \item Calculate the coupling terms (jump conditions) for the discontinuous derivatives to arrive at the matched initial value system, which yields a non-trivial solution characterized by a real-valued growth rate.
        \end{enumerate}
\end{enumerate}

\section{Mathematical Formulation}\label{sec:mathformulation}
In this section, we develop the mathematical formulation for two-dimensional flow in an isotropic porous medium with uniform properties, for a two-phase, two-component system in thermodynamic equilibrium. We identify the components in our binary mixture as ``$a$'' (the lighter component) and ``$b$'' (the heavier component). 

We are interested in the case of a gas (rich in component $a$) displacing an undersaturated liquid (rich in component $b$). Flow is considered uniform in the \mbox{$x$-direction} with a shock at the interface between the invading and displaced fluids. We assume that if instability occurs, it will be at the shock where the total mobility gradient is highest---which \citet{riaz2004} shows is a reasonable assumption. Therefore, after defining our \textit{base-state} solution as the one-dimensional, travelling-wave solution around the shock, our analysis then consists in subjecting the base-state to two-dimensional, small wave-like perturbations, and checking the stability of the system.

\begin{figure}[!htb]
    \centering
    \includegraphics[width=1\linewidth]{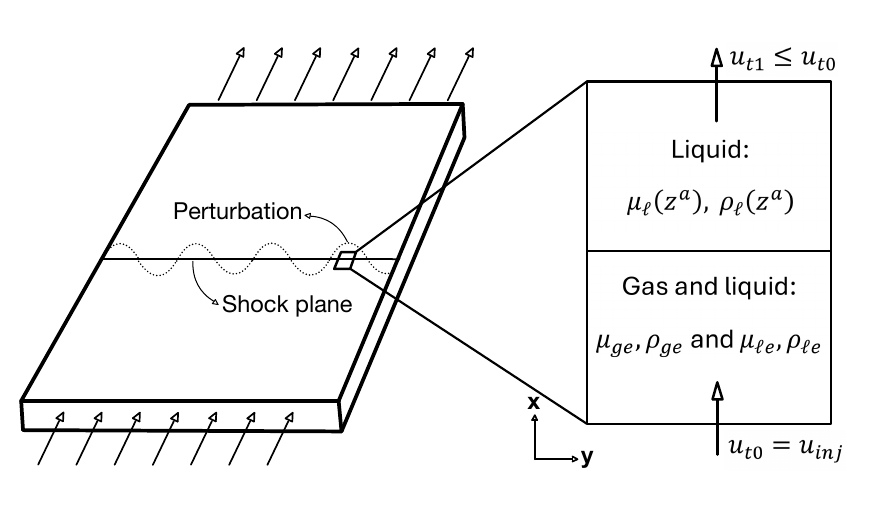}
    \caption{Flow diagram of the gas-displacing-liquid process in the two-component case. Upstream of the shock front, gas and liquid phases are present, with their properties as functions of their equilibrium compositions. Downstream of the shock, the only phase is an undersaturated fluid, and total flow velocity ($u_{t1}$) is (generally) lower than the upstream total velocity ($u_{t0}$).}\label{fig:flowdiagram}
\end{figure}

Figure~\ref{fig:flowdiagram} shows a sketch of the system configuration. Although not explicitly shown in our diagram, gravity is accounted for in our formulations and results shown in Section~\ref{sec:results} encompass three displacement directions: horizontal, upward and downward.

\begin{figure}[!htb]
    \centering
    \includegraphics[width=1\linewidth]{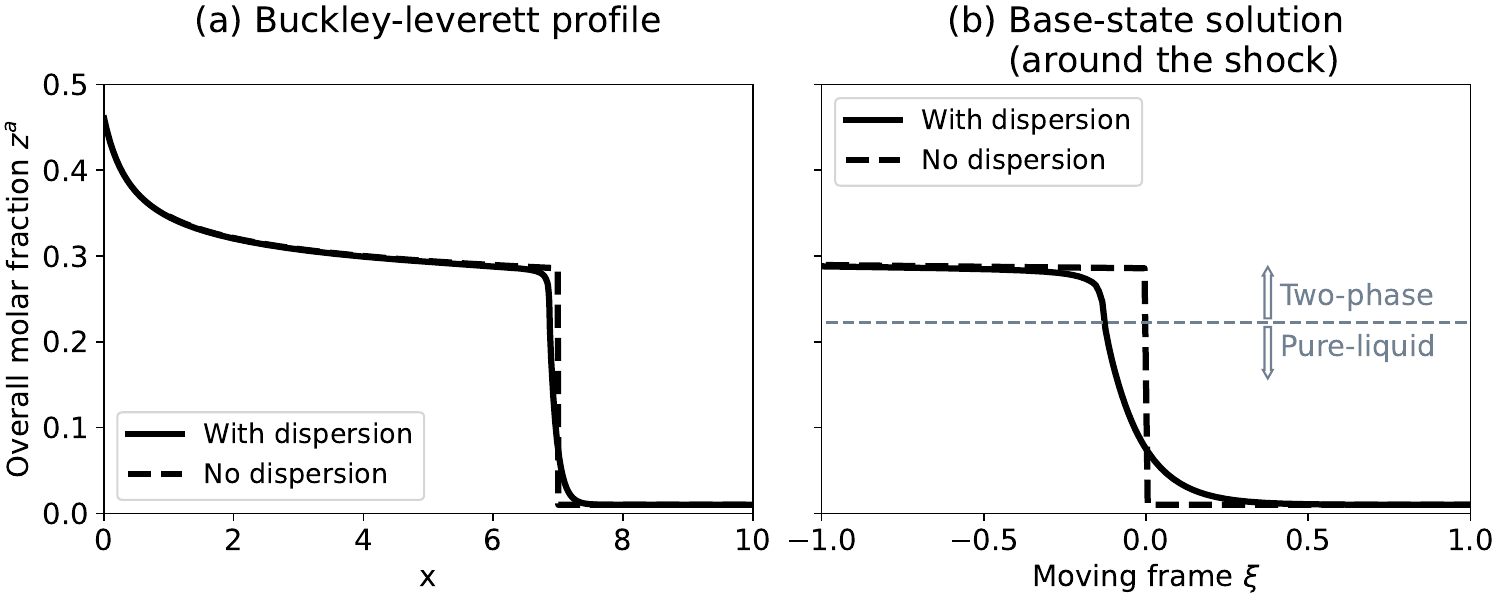}
    \caption{Buckley-Leverett solution and the base-state solution (travelling-wave solution around the shock at ${\xi=0}$, where ${\xi=x-v_s\,t}$). The case with no dispersion (no capillary force, molecular diffusion or mechanical dispersion) is shown for comparison.}\label{fig:base_vs_BL}
\end{figure}

Figure~\ref{fig:base_vs_BL} shows an example of a \textit{full} Buckley-Leverett profile for the overall composition of component $a$ ($z^a$), and the solution at a localized region around the shock, which at large displacement periods converges to our base-state solution.

In the following subsections, we first discuss the \textit{flash} equations for a system of only two components, list the simplifying assumptions, write the two main flow equations, briefly discuss the calculation of the shock in the presence of mass transfer, develop the \textit{base-state} equations given by the moving-frame transformation, and, finally, we derive the perturbation equations and the coupling solution around the phase-change discontinuity.

\FloatBarrier\subsection{Phase equilibrium}\label{sec:phase_eq}
Phase equilibrium calculations are considerably simpler for binary systems than for systems containing three or more components. To illustrate this, we first consider a mixture of $N_c$ components with its overall composition given by ${z^\ga=n^\ga/n_t}$, ${\ga=1,\ldots,N_c}$, with $n^\ga$ as the number of moles of component $\ga$ and $n_t$ as the total number of moles in the system. Suppose that at a certain pressure and temperature ($P,T$) this system \textit{splits} into two phases, gas and liquid, with their respective compositions given by ${y^\ga=n^\ga_g/n_g}$ and ${x^\ga=n^\ga_\ell/n_\ell}$, where $n_\al$ is the number of moles of phase $\al$ and $n^\ga_\al$ the number of moles of $\ga$ in $\al$. The two phases are in thermodynamic equilibrium if the fugacity of each component is the same in both phases \citep{pedersen2006}:
\begin{equation}
    \hat{f}_g^\ga(P, T, \{y\}) = \hat{f}_\ell^\ga(P, T, \{x\}),\quad \text{for}\quad \ga=1,\ldots,N_c,
\label{eq:fugacity}
\end{equation}
where $\hat{f}_\al^\ga$ is the fugacity of component $\ga$ in phase $\al$, ${\{y\}=\{y^1,\ldots,y^{N_c-1}\}}$ and ${\{x\}=\{x^1,\ldots,x^{N_c-1}\}}$ are the sets of independent values of the gas and liquid compositions (since, by definition, ${\sum{y^\ga}=1}$ and ${\sum{x^\ga}=1}$). The expressions for $\hat{f}_\al^\ga$ can be written with an appropriate equation of state (EOS). 

Assuming the system is in equilibrium, the composition of the gas and liquid phases can be found by a \textit{flash} calculation, which consists of solving the thermodynamic equilibrium equations (\ref{eq:fugacity}) together with the material balance, or \textit{tie-line}, equations that link the overall composition of the mixture with the equilibrium compositions:
\begin{equation}
    z^\ga = L_g\,y^\ga + (1-L_g)\,x^\ga,\quad \text{for}\quad \ga=1,\ldots,N_c-1,
\label{eq:tieline}
\end{equation}
where ${L_g=n_g/n_t}$ is the gas mole fraction.

In total, there are $2N_c-1$ primary unknowns and $2N_c-1$ independent equations, so the system is perfectly determined. If $N_c\geq3$, Equations~\ref{eq:fugacity} and~\ref{eq:tieline} need to be solved simultaneously. However, for the binary case (${N_c=2}$), we have two unknowns and two equations in Equation~\ref{eq:fugacity}, and the equilibrium compositions can be solved independently of the overall composition---i.e. ${y^\ga=y^\ga(P,T)}$ and ${x^\ga=x^\ga(P,T)}$. Equation~\ref{eq:tieline} is then decoupled from Equation~\ref{eq:fugacity}, and with $z^\ga$, $y^\ga$ and $x^\ga$ given, we can solve for $L_g$:
\begin{equation}
    L_g = \frac{z^\ga - x^\ga_e}{y^\ga_e - x^\ga_e},\quad \text{for}\quad \ga=a,b,
\label{eq:tieline2}
\end{equation}
where $y_e^\ga$ and $x_e^\ga$ are the \textit{constant} equilibrium compositions of the gas and liquid phases inside the two-phase region.

In summary, we need to solve Equation~\ref{eq:fugacity} only once to obtain the equilibrium compositions of the gas and liquid phases in the two-phase region, and the \textit{flash} calculation is reduced to solving Equation~\ref{eq:tieline2} for the overall composition $z^\ga$. We can also easily identify the two-phase and single-phase regions. Indeed, assuming that component $a$ is \textit{lighter} than component $b$ (i.e. $y_e^a > x_e^a$ and $y_e^b < x_e^b$), \citet{orr2007} makes the following observations regarding Equation~\ref{eq:tieline2}:
\begin{enumerate}
    \item If $z^a > y_e^a$, then the mixture is an undersaturated (or single-phase) gas, and also $L_g > 1$;
    \item If $z^a < x_e^a$, the mixture is an undersaturated (or single-phase) liquid, and $L_g < 0$;
    \item $y_e^\ga$ and $x_e^\ga$ are constant inside the two-phase region ($x_e^a \leq z^a \leq y_e^a$).
\end{enumerate}

Item (3) implies that, for a fixed ($P,T$), the properties---such as viscosity ($\mu_\al$) and density ($\rho_\al$)---for the gas and liquid phases are constant in two-phase region. In the single-phase regions, that is usually not true, and $\mu_\al$ and $\rho_\al$ are dependent of the overall composition. However, the density of the liquid phase may only be weakly dependent on $z^a$, and for simplicity we assume that $\rho_\ell=constant$.

Normally, to calculate the phase density we calculate its molar volume $V_\al$ using our EOS, and then use the relation:
\begin{equation}
    \rho_\al = \bar{M}_\al c_\al,
\label{eq:density1}
\end{equation}
where $\bar{M}_\al$ is the average molar weight of phase $\al$ and $c_\al=1/V_\al$ its molar concentration (or molar density).

For the binary case ($\ga=a,b$), in the pure-liquid region, we can write:
\begin{equation}
    \bar{M}_\ell = \sum_\ga M^\ga z^\ga = M^b + (M^a - M^b) z^a.
\label{eq:Mwmix}
\end{equation}

With $\rho_{\ell e}$ the value of the liquid density in the two-phase region, we can ensure that $\rho_\ell = \rho_{\ell e}$ in the single-phase liquid region by calculating $c_\ell$ with Equation~\ref{eq:density2}, instead of using the EOS\@.

\begin{equation}
    c_\ell = \frac{\bar{M}_\ell}{\rho_{\ell e}}.
\label{eq:density2}
\end{equation}

\subsection{Flow equations}
For the derivation of the flow equations in the next sections, we make the following simplifying assumptions:
\begin{enumerate}
    \item The fluid injected is in the two-phase region ($y^a < z^a < x^a$), and the single-phase gas region is bypassed;
    \item The liquid phase density in the single-phase liquid region is independent of the liquid composition and equal to $\rho_{\ell e}$;
    \item The temperature is constant;
    \item The pressure drop along flow is small compared to the initial pressure such that the fluid properties are evaluated at initial pressure;
    \item The porous medium is uniform and isotropic;
    \item Mechanical dispersion dominates over molecular diffusion; can be described by a Fickian-type equation; and the dispersivities are independent of the molecular concentrations (more details in Section~\ref{sec:dispflux});
    \item The capillary pressure derivative with respect to gas saturation is a function of the phases' relative permeabilities (see Section~\ref{sec:dPc}).
\end{enumerate}

\subsubsection{Mass balance}
The material balance equation of component $\ga$ in phase $\al$, for a multiphase, compositional flow in porous media, for a bounded domain $\Omega \subset \mathbb{R}^n\, (n=1,2)$, without the presence of source/sink terms, is given by~\citep{bear2018}:
\begin{equation}
    \phi \frac{\p S_\al \rho_\al^\ga}{\p t} + \phi \nabla \cdot \left[S_\al \left(\rho_\al^\ga \bol{V}_\al + \bol{J}_\al^{\ga}\right) \right] - \sum_{\beta\neq\al} f_{\beta\to\al}^\ga = 0,
\label{eq:matbal}
\end{equation}
where $\phi$ is the medium porosity, $S_{\al}$ the saturation of phase $\al$, $\rho_\al^\ga$ the mass density of component $\ga$ in phase $\al$, $\bol{V}_\al$ is the phase velocity, $\bol{J}_\al^{\ga}$ represents the sum of diffusive and dispersive fluxes of component $\ga$ in phase $\al$ and $f_{\beta\to\al}^\ga$ represents the transport of $\ga$-mass from phase $\beta$ to phase $\al$.

Recognizing that in a two-phase system $f_{\beta\to\al}^\ga=-f_{\al\to\beta}^\ga$, we arrive at the equation for the total mass balance of $\ga$ by summing over phases $g$ (gas) and $l$ (liquid) in Equation~\ref{eq:matbal}:
\begin{equation}
    \phi \frac{\p \left(S_g \rho_g^\ga + S_\ell \rho_\ell^\ga \right) }{\p t} + \nabla \cdot \left(\rho_g^\ga\bol{u}_g + \rho_\ell^\ga\bol{u}_\ell \right) + \phi \nabla \cdot \left[ \left( S_g \bol{J}_g^{\ga} + S_\ell \bol{J}_\ell^{\ga} \right) \right] = 0,
\label{eq:matbal_total1}
\end{equation}
where $\bol{u}_\al=\phi S_\al \bol{V}_\al$ is the specific discharge---or Darcy velocity---of phase $\al$.

\subsubsection{Dispersive flux}\label{sec:dispflux}
For simplicity, we assume that mechanical dispersion dominates over molecular diffusion. For the case of uniform flow in an isotropic medium, we can express the mechanical dispersion flux in a \textit{Fickian-type} law as \citep{bear2018}:
\begin{equation}
    \bol{J}_{\al}^\ga = -M^\ga \bol{D}_{\al}^{\ga\delta} \nabla c_\al^\ga,
\label{eq:fickdisp}
\end{equation}
where $\bol{D}_{\al}^{\ga\delta}$ is the dispersion tensor given by:
\begin{equation}
    \bol{D}_{\al}^{\ga\delta} = 
    \begin{bmatrix}
        a_L & 0 \\
        0 & a_T \\
    \end{bmatrix} V_{x,\al},
\label{eq:disptensor}
\end{equation}
where $a_L$ and $a_T$ are the longitudinal and transverse dispersivities.

We assume the dispersivities are independent of $c_\al^\ga$, and so the condition ${\bol{J}_\al^\ga + \bol{J}_\al^\delta = 0}$ implies that $\bol{D}_{\al}^{\ga\delta}=\bol{D}_{\al}^{\delta\ga}=\bol{D}_{\al}$. The total dispersion flux becomes:
\begin{equation}
    \bol{J}_{\al}^\ga = -M^\ga \bol{D}_{\al} \nabla c_\al^\ga,\quad\text{with}\quad
    \bol{D}_{\al} =
    \begin{bmatrix}
        a_L & 0 \\
        0 & a_T \\
    \end{bmatrix} \frac{u_{x,\al}}{\phi}.
\label{eq:ficktotal}
\end{equation}

In the two-phase region, $\nabla c_\al^\ga=\bol{0}$, so the dispersion flux term is only non-zero at the single-phase regions. Furthermore, since we assumed that the injected fluid is in the two-phase region, the dispersion in the gas phase is inconsequential. We can then write the material balance equation (\ref{eq:matbal_total1}) as:
\begin{equation}
    \phi \frac{\p \left(S_g c_g^\ga + S_\ell c_\ell^\ga \right) }{\p t} + \nabla \cdot \left(c_g^\ga\bol{u}_g + c_\ell^\ga\bol{u}_\ell \right) - \phi \nabla \cdot \left(\bol{D}_{\ell} \nabla c_\ell^\ga \right) = 0,
\label{eq:matbal_total2}
\end{equation}

\subsubsection{Darcy's law}
For multiphase flow in an isotropic porous media, Darcy's law can be written as:

\begin{equation}
    \bol{u}_\al = \la_\al \left(\nabla P_\al + \rho_\al \bol{g} \right),
\label{eq:darcy}
\end{equation}
where $P_\al$ is the pressure and $\rho_\al$ the mass density of phase $\al$, $\bol{g}=g\bol{e}_x$ is the acceleration vector due to gravity, $\bol{e}_x$ being the unit vector in the vertical direction $x$, and $\la_\al$ the mobility function given by:
\begin{equation}
    \la_\al = \frac{k_{r\al}}{\mu_\al} k,
\label{eq:lambda}
\end{equation}
where $k$ is the absolute permeability, $k_{r\al}$ is the relative permeability function and $\mu_\al$ is the viscosity of phase $\al$.

The total Darcy velocity $\bol{u}_t=\bol{u}_g + \bol{u}_\ell$ can be written as:
\begin{equation}
    \mathbf{u}_t = -\lambda_t \nabla P_\ell - \lambda_g \nabla P_c - (\lambda_g \rho_g + \lambda_\ell \rho_\ell) \mathbf{g},
\label{eq:ut}
\end{equation}
where $\la_t=\la_g+\la_\ell$ and $P_c$ is the capillary pressure given by:
\begin{align}
    P_c = P_g-P_\ell = \sqrt{\frac{\phi}{k}}\ \sig_{g\ell}\,\cos\theta_{g\ell}\, J_c,
\label{eq:Pc}
\end{align}
where $\sig_{g\ell}$ is the gas-liquid interfacial tension, $\theta_{g\ell}$ is the contact angle and $J_c$ is the Leverett $J$-function.

We can write the phase velocities as functions of the total velocity:
\begin{equation}
\begin{aligned}
    \bol{u}_g &= f_g \mathbf{u} - \lambda \nabla P_c + \lambda \Delta\rho \mathbf{g}, \\
    \bol{u}_\ell &= f_\ell \mathbf{u} + \lambda \nabla P_c - \lambda \Delta\rho \mathbf{g},
\end{aligned}
\label{eq:phasevelo}
\end{equation}
where $f_\al=\lambda_\al/\lambda_t$ is the fractional flow of phase $\al$, $\lambda=\lambda_g\lambda_\ell/\lambda_t$ and $\Delta\rho=\rho_\ell-\rho_g$.

\subsubsection{Mass balance in terms of total velocity}
By substituting Equation~\ref{eq:phasevelo} in Equation~\ref{eq:matbal_total2}, we arrive at the material balance equation in terms of the total Darcy velocity.

\begin{equation}
    \phi \frac{\p C^\ga }{\p t} + \nabla \cdot F^\ga - \nabla \cdot \left( \bol{D}_t^\ga \nabla C^\ga \right) = 0,\quad \left\{ \begin{aligned}
        &C^\ga = S_g c_g^\ga + S_\ell c_\ell^\ga, \\
        &F^\ga = \left( f_g c_g^\ga + f_\ell c_\ell^\ga \right) \bol{u}_t + \lambda\Delta\rho \Delta c^\ga \bol{g}, \\
        &\bol{D}_t^\ga = \lambda\Delta c^\ga \frac{dP_c}{dC^\ga}\bol{I} + \phi \delta_\ell \bol{D}_{\ell}, \\
        &\Delta c^\ga = c_g^\ga-c_\ell^\ga,
        \end{aligned}\right.
\label{eq:matbal_ut}
\end{equation}
where $C^\ga$ is the overall concentration, $F^\ga$ is the total flow function and $D_t^\ga$ the total dissipative tensor for component $\ga$. The term $\delta_\ell$ is the \textit{indicator function} for the single-phase liquid (or \textit{pure-liquid}) region defined by:
\begin{equation}
    \delta_\ell = 
    \left\{ \begin{aligned}
        & 0,\quad z^b < x^b_e,\\
        & 1,\quad z^b \geq x^b_e.
    \end{aligned} \right.
\end{equation}

In the one-dimensional, incompressible case, the total velocity becomes a known constant, and so the one-dimensional version of Equation~\ref{eq:matbal_ut} becomes quite useful as the base-state equation.

\subsubsection{Mass balance in terms of phase pressure}
For the two-dimensional case, $\bol{u}_t$ is not known \textit{a priori}, and it is preferable to write the mass balance equation in terms of the overall concentration and the pressure. By substituting Equation~\ref{eq:darcy} in Equation~\ref{eq:matbal_total2}, we arrive at the material balance equation in terms of the total liquid phase pressure.

\begin{equation}
    \phi \frac{\partial C^\ga}{\partial t} - \nabla \cdot \left( \bar{\lambda}^\ga \nabla P_\ell + \bar{\lambda}_\rho^\ga \bol{g} \right) - \nabla \cdot \left(\bsym{\Lambda}_t^\ga \nabla C^\ga \right) = 0,\quad \left\{ \begin{aligned}
        &\bar{\lambda}^\ga = \lambda_g c_g^\ga + \lambda_\ell c_\ell^\ga, \\
        &\bar{\lambda}_\rho^\ga = \lambda_g c_g^\ga \rho_g^\ga + \lambda_\ell c_\ell^\ga \rho_\ell^\ga, \\
        &\bsym{\Lambda}_t^\ga = \lambda_g c_g^\ga \frac{dP_c}{dC^\ga} \bol{I} + \phi \delta_\ell \bol{D}_\ell.
        \end{aligned}\right.
\label{eq:matbal_pl}
\end{equation}

\subsubsection{Total velocity}\label{sec:total_velocity}
The assumptions made in Section~\ref{sec:phase_eq} mean that fluid properties are constant inside the two-phase region, and that liquid density is also constant in the pure-liquid region. That leads to the useful result ($\nabla \cdot \bol{u}_t =0$). In the two-phase region, this can be proven by rewriting Equation~\ref{eq:matbal_ut} with the knowledge that $c_\al^\ga$ is constant in the two-phase region, to obtain:

\begin{equation}
    \phi\frac{\p S_g}{\p t} + \nabla \cdot \left[ 
    \bol{u}_t \left( f_g + \frac{c_\ell^\ga}{\Delta c^\ga} \right) + \lambda \Delta\rho \bol{g} + \lambda \nabla P_c\right] = 0.
\label{eq:matbal_sg}
\end{equation}

By subtracting Equation~\ref{eq:matbal_sg} with $\ga=b$ from Equation~\ref{eq:matbal_sg} with $\ga=a$, we arrive at:

\begin{equation}
    \left( \frac{c_\ell^a}{\Delta c^a} - \frac{c_\ell^b}{\Delta c^b} \right) \nabla \cdot \bol{u}_t = 0\quad \implies \quad \nabla \cdot \bol{u}_t=0.
\end{equation}

For the pure-liquid region, where ${\bol{u}_t=\bol{u}_\ell}$, and assuming ${\rho_\ell=constant}$, Equation~\ref{eq:matbal_total2} becomes:
\begin{equation}
    \phi \frac{\p \rho_\ell^\ga}{\p t} + \nabla \cdot \left( \rho_\ell^\ga \bol{u}_t \right) + \phi \nabla \cdot \bol{J}_\ell^\ga = 0,
\label{eq:const_u_proof2}
\end{equation}
and by summing over $\ga$, and recognizing that $\rho_\ell^a + \rho_\ell^b =\rho_\ell$ and $\bol{J}_\ell^a + \bol{J}_\ell^b = 0$, we arrive at:
\begin{equation}
    \rho_\ell \nabla \cdot \bol{u}_t = 0.
\end{equation}

With the result $\nabla\cdot\bol{u}_t=0$ holding for both the two-phase and pure-liquid regions, we can then write:
\begin{equation}
    -\nabla \cdot \bol{u}_t = \nabla \cdot \left( \lambda_t \nabla P_\ell + \lambda_g \nabla P_c + \lambda_\rho \mathbf{g} \right) = 0,\quad \text{with}\quad \lambda_\rho = \lambda_g \rho_g + \lambda_\ell \rho_\ell.
\label{eq:gradut}
\end{equation}

Equations~\ref{eq:matbal_pl} and~\ref{eq:gradut} form the system of equations for the two-dimensional, \textit{perturbed} flow case, and can be solved for $C^\ga$ and $P_\ell$.

Finally, because the term $c_\ell^\ga/\Delta c^\ga$ is a constant, for the two-phase region, Equation~\ref{eq:matbal_sg} can be rewritten as:
\begin{equation}
    \phi\frac{\p S_g}{\p t} + \nabla \cdot \left( f_g \bol{u}_t + \lambda \Delta\rho \bol{g} + \lambda \nabla P_c \right) = 0,
\label{eq:matbal_sg_2}
\end{equation}
which is the well-known balance equation for incompressible, immiscible two-phase flow. This means that, in the case where the initial fluid to be displaced is a saturated liquid, and therefore flow is restricted entirely in the two-phase region, our two-phase, two-component model reduces to the incompressible, immiscible two-phase model.

\subsection{Non-dimensionalization}
We can write our equations in dimensionless by the following normalizations:
\begin{equation}
    \begin{aligned}
        \bol{x}^*&=\frac{\bol{x}}{L}, &\bol{u}^*=&\frac{\bol{u}}{u_{inj}}, &t^*=&\frac{u_{inj}}{\phi L}t,\\
        c_\al^{\ga*}&=\frac{c_\al^\ga}{c_{\ell e}}, &\mu_{\al*}=&\frac{\mu_\al}{\mu_{g e}}, &\lambda_\al^*=&\frac{\mu_{g e}}{k}\lambda_\al = \frac{k_{r\al}}{\mu_\al^*},\\
        P_\al^*&=\frac{k}{u_{inj}\,\mu_{ge}\, L}P_\al, &\rho_\al^*=&\frac{kg}{u_{inj}\,\mu_{g e}}\rho_\al, &\bol{D}_\al^*=&\frac{\phi}{u_{inj}\,L} \bol{D}_\al,
    \end{aligned}
\label{eq:normalization}        
\end{equation}
where $u_{inj}$ is the injection specific discharge, and $c_{\ell e}$ and $\mu_{ge}$ are the molar concentration of the liquid phase and the viscosity of the gas phase, respectively, inside the two-phase region. The length scale $L$ is chosen as:
\begin{equation}
    L = \frac{\sqrt{\phi\, k}}{N_{ca}},\quad N_{ca} = \frac{u_{inj}\, \mu_{ge}}{A_{pc}\,\sig_{g\ell}\,\cos\theta_{g\ell}},
\label{eq:lenghtscale}
\end{equation}
where $N_{ca}$ is the capillary number and $A_{pc}$ is a multiplying constant in the $J_c$ function. 

With the normalizations in~\ref{eq:normalization}, and dropping the superscript ``*'', Equations~\ref{eq:matbal_ut},~\ref{eq:matbal_pl} and~\ref{eq:ut} can be rewritten as:
\begin{align}
    &\frac{\p C^\ga}{\p t} + \nabla \cdot F^\ga - \nabla \cdot \left( \bol{D}_t^\ga \nabla C^\ga \right) = 0,\quad \left\{ \begin{aligned}
        &C^\ga = S_g c_g^\ga + S_\ell c_\ell^\ga, \\
        &F^\ga = \left( f_g c_g^\ga + f_\ell c_\ell^\ga \right) \bol{u}_t + \lambda\Delta\rho \Delta c^\ga \bol{e}_g, \\
        &\bol{D}_t^\ga = \lambda\Delta c^\ga \frac{dP_c}{dC^\ga} \bol{I} + \delta_\ell\bol{D}_{\ell}, \\
        &\Delta c^\ga = c_g^\ga-c_\ell^\ga,
        \end{aligned}\right.
    \label{eq:matbal_ut_nond} \\
    &\frac{\partial C^\ga}{\partial t} - \nabla \cdot \left( \bar{\lambda}^\ga \nabla P_\ell + \bar{\lambda}_\rho^\ga \bol{e}_g \right) - \nabla \cdot \left(\bsym{\Lambda}_t^\ga \nabla C^\ga \right) = 0,\quad \left\{ \begin{aligned}
        &\bar{\lambda}^\ga = \lambda_g c_g^\ga + \lambda_\ell c_\ell^\ga, \\
        &\bar{\lambda}_\rho^\ga = \lambda_g c_g^\ga \rho_g^\ga + \lambda_\ell c_\ell^\ga \rho_\ell^\ga, \\
        &\bsym{\Lambda}_t^\ga = \lambda_g c_g^\ga \frac{dP_c}{dC^\ga} + \delta_\ell\bol{D}_\ell,
        \end{aligned}\right.
    \label{eq:matbal_pl_nond} \\
    &\bol{u}_t = \lambda_t \nabla P_\ell + \lambda_\rho \mathbf{e}_g + \lambda_c \nabla C^\ga,\quad \left\{ \begin{aligned} 
        & \lambda_c = \lambda_g \frac{dP_c}{dC^\ga}, \\
        & \lambda_\rho = \lambda_g \rho_g + \lambda_\ell \rho_\ell,
    \end{aligned}\right.
    \label{eq:ut_nond}
\end{align}
where $\mathbf{e}_g$ is the unit vector of gravity.

Henceforth, all equations are assumed to be in dimensionless form, unless indicated otherwise.

\subsection{Shock in a two-component system}\label{sec:shock}
In Section~\ref{sec:mathformulation}, we briefly introduced the base-state solution as the one-dimensional, travelling-wave solution (to Equation~\ref{eq:matbal_ut}) that follows the shock between the injected and displaced fluids. This solution, naturally, requires knowledge of the shock front properties, particularly the shock velocity $v_s$. However, when mass-transfer effects are present, calculating the shock conditions becomes more involved than in the classic immiscible case. \citet{orr2007} has demonstrated how to calculate the flow profile for a two-phase, two-component problem using the \textit{method of characteristics}. In this section, we provide a brief overview of this procedure. 

For the two-component case, and inside the two-phase region where the equilibrium concentrations are constants, one can write:
\begin{equation}
    S_g = \frac{C^\ga - c_\ell^\ga}{c_g^\ga - c_\ell^\ga}.
\label{eq:tieline_sat}
\end{equation}

Equation~\ref{eq:tieline_sat} shows that phase saturation is a function only of the overall concentration $C^\ga$, and therefore we can write that ${F^\ga=F^\ga(C^\ga, S_g(C^\ga))=F^\ga(C^\ga)}$. By ignoring the dissipative terms, we write the purely advective, one-dimensional version of Equation~\ref{eq:matbal_ut_nond} as:
\begin{equation}
    \frac{\p C^\ga}{\p t} + \frac{d F^\ga}{d C^\ga} \frac{dC^\ga}{d x} = 0.
\end{equation}

Along a characteristic curve $\left(x(\eta),t(\eta)\right)$ emanating from a starting position $x_0$, the characteristic equations are given by:
\begin{equation}
    \begin{aligned}
        \frac{dt}{d\eta} = 1,\quad \frac{dx}{d\eta}=\frac{dF^\ga}{dC^\ga},\quad \frac{dC^\ga}{d\eta}=0.
    \end{aligned}
\end{equation}

So the overall concentration $C^\ga$ is constant along the characteristic curves, which in this case are straight lines given by:
\begin{equation}
    x = \frac{dF^\ga}{dC^\ga}\,t + x_{t0},
\end{equation}
where $x_{t0}$ is the initial position at which $dF^\ga/dC^\ga$ is evaluated.

We assume that the initial system is homogeneous so that, at $x_{t0}>0$, $C^\ga(x)=C^\ga_i$, $C^\ga_i$ being the initial overall concentration of $\ga$. At the origin $x_{t0}=0$, $C^\ga=\left[C^\ga_i, C^\ga_{inj}\right]$, where $C^\ga_{inj}$ is the overall concentration of the injected fluid, and an infinite number of characteristic lines emanate from it. Not all characteristic lines are physically possible, however, and \textit{shocks} will occur when characteristic lines cross-over each other.

\citet{orr2007} shows that for the two-phase, two-component system, where the initial in-situ fluid is a pure liquid, and the injected fluid is a pure gas, two shocks will appear: a \textit{leading shock} at the transition from the pure-liquid to the two-phase region, and a \textit{trailing shock} at the transition from the two-phase to the pure-gas region. However, we are only interested in the leading shock, as the range for $C^\ga$ in the solution of the base-state equation (see Section~\ref{sec:basestate}, Equation~\ref{eq:basestate_C}) is $\left[C^\ga_{s}, C^\ga_0\right]$, with $C^\ga_s$ the leading shock overall concentration in the two-phase region. To simplify matters, we assume that our injected fluid is a two-phase fluid, and therefore only the leading shock needs to be considered.

The solution for the shock ($C^\ga=C^\ga_s$) must satisfy three conditions \citep{orr2007}:
\begin{enumerate}
    \item \textit{Jump condition}: the Ranking-Hugoniot relation, given by
    \begin{equation}
        v_s = \left. \frac{dx}{dt} \right|_s = \frac{F^\ga_0 - F^\ga_1}{C^\ga_0 - C^\ga_1},
    \label{eq:jump1}
    \end{equation}
    where the subscripts 0 and 1 indicates the upstream and downstream sides of the shock, respectively;
    \item \textit{Entropy condition}:
    \begin{equation}
        \left. \frac{dx}{dt} \right|_1 \leq \left. \frac{dx}{dt} \right|_s \leq \left. \frac{dx}{dt} \right|_0;
    \end{equation}
    \item \textit{Velocity constraint}: plane wave velocities must decrease monotonically for the solution from the shock composition to the injection composition---i.e. $v_s$ must be maximal.
\end{enumerate}

Conditions (2) and (3) imply that:
\begin{equation}
    \left. \frac{dx}{dt} \right|_0 = v_s,\quad \left. \frac{dx}{dt} \right|_1 < v_s,
\end{equation}
and conditions (1), (2) and (3) imply that:
\begin{equation}
    \left. \frac{dF^\ga}{dC^\ga} \right|_s = \left. \frac{dF^\ga}{dC^\ga} \right|_0 = \frac{F^\ga_0 - F^\ga_1}{C^\ga_0 - C^\ga_1}.
\label{eq:jump2}
\end{equation}

We assume that the shock occurs at the two-phase region, where we can write:
\begin{equation}
    \left. \frac{dF^\ga}{dC^\ga} \right|_0 = \left. \frac{dF^\ga}{dS_g} \frac{dS_g}{dC^\ga} \right|_0 = u_t \left. \frac{df_g}{dS_g} \right|_0 + \Delta\rho \left. \frac{d\lambda}{dS_g} \right|_0,
\label{eq:2p_dFdC}
\end{equation}
and the jump condition becomes:
\begin{equation}
    u_t \left. \frac{df_g}{dS_g} \right|_0 + \Delta\rho \left. \frac{d\lambda}{dS_g} \right|_0 =  \frac{F^\ga_0 - F^\ga_1}{C^\ga_0 - C^\ga_1}.
\label{eq:jump3}
\end{equation}

We need to evaluate the function $u_t(C^\ga)$ at the upstream and downstream compositions to evaluate $F^\ga_0$ and $F^\ga_1$. As was established earlier, $u_t$ is constant in both the two-phase and pure-liquid regions. However, if the components undergo a change in volume when they are transferred from one phase to another, the total velocity is not conserved along the shock, meaning that $u_t(C^\ga_0)\neq u_t(C^\ga_1)$. Since we assumed the injected fluid to be in the two-phase region, we simply have $u_{t0}=u_t(C^\ga_0)=u_{inj}$. The value for $u_{t1}=u_t(C^\ga_1)$ remains undetermined, and therefore an extra equation is needed, and is provided by the fact that Equation~\ref{eq:jump1} is independently true for both $\ga=a$ and $\ga=b$, which allows us to write:
\begin{equation}
    \frac{F^a_0 - F^a_1}{C^a_0 - C^a_1} = \frac{F^b_0 - F^b_1}{C^b_0 - C^b_1}.
\label{eq:shock_2ndcond}
\end{equation}

Finally, we can solve for ($C^\ga_0,u_{t1}$) using Equations~\ref{eq:jump3} and~\ref{eq:shock_2ndcond}.

\subsection{The base-state equation}\label{sec:basestate}
From Reynold's theorem, the material balance of an extensive quantity $E$, with density $e$, in a moving-frame of reference $\bol{\xi} = \bol{x} - \bol{v} t$, with $\bol{V}^E=\p\bol{x}/\p t$ as the material element velocity, can be written as:
\begin{equation}
    \begin{aligned}
        \int_{\Omega_{E(t)}} e\, \text{d}\volume &= \int_{\Omega_{E(t)}} \frac{\p e}{\p t} + \frac{\partial}{\partial x_i} \left(e \frac{\p x_i}{\p t}\right) \, \text{d}\volume\\
        &= \int_{\Omega_{E(t)}} \frac{\p e}{\p t} - v_i\frac{\partial e}{\partial \xi_i} + \frac{\partial}{\partial \xi_i} (e V^E_i) \, \text{d}\volume,
    \end{aligned}
\end{equation}
and the conservation differential equation for an arbitrary elementary volume $\Omega_{E(t)}$ becomes:
\begin{equation}
    \frac{\p e}{\p t} - \bol{v} \cdot \nabla_{\xi}\, e + \nabla_\xi \cdot \left( e \bol{V}^E \right) = 0
\label{eq:matbal_emf}
\end{equation}

From Equation~\ref{eq:matbal_emf}, we observe that the moving-frame equation---in terms of $(\xi,t)$---is obtained simply by subtracting the term ${\bol{v}\cdot\nabla_{\xi}e}$ from the static-frame equation. So for the one-dimensional case of uniform flow, the moving-frame formulation to Equation~\ref{eq:matbal_ut_nond} (where ${e=C^\ga}$) that follows the shock front is:
\begin{equation}
    \frac{\p C^\ga}{\p t} - v_s \frac{\p C^\ga}{\p \xi} + \frac{\p F^\ga}{\p \xi} - \frac{\p}{\p \xi} \left( D_{t,xx}^\ga \frac{\p z^\ga}{\p \xi}\right) = 0.
\label{eq:matbal_ut_mf}
\end{equation}

We could numerically solve Equation~\ref{eq:matbal_ut_mf} to obtain a time-dependent, Buckley-Leverett-type solution for the overall concentration $C^\ga$. By assuming that perturbations grow or decay faster than such solution changes with time (the quasi-steady-state assumption), the base state would then be set for a given time $t$. We opt, however, to use the steady-state version of Equation~\ref{eq:matbal_ut_mf} (where for ${t\to\infty}$, ${\p C^\ga/\p t \to 0}$) as our base-state equation.

We can rearrange Equation~\ref{eq:matbal_ut_mf} and integrate to obtain:
\begin{equation}
    \int_{\xi_0}^\xi \frac{d}{d \xi} \left( D_{t,xx}^\ga \frac{d C^\ga}{d \xi}\right) d\xi = \int_{\xi_0}^\xi \frac{d}{d \xi} \left(v_s C^\ga - F^\ga\right) d\xi,
\label{eq:matbal_ut_mf_ss}
\end{equation}
where $D_{t,xx}^\ga$ is the diagonal term in $x$-direction of tensor $\bol{D}_t^\ga$.

In the steady-state case, the transition zone around the shock is infinitely large, with $C^\ga=C^\ga_i$ ($C^\ga_i$ being the initial overall concentration of $\ga$) and ${\p_\xi C^\ga=0}$ when $\xi_0\to \infty$, and the \textit{base-state equation} is simplified to:
\begin{equation}
    D_{t,xx}^\ga \frac{d C^\ga}{d \xi} = F_i^\ga - F^\ga + v_s \left( C^\ga - C_i^\ga \right),
\label{eq:basestate_C}
\end{equation}
where $F_i^\ga = F^\ga(z_i^\ga)$ and $C_i^\ga = C^\ga(z_i^\ga)$.

We observe that the solution to Equation~\ref{eq:basestate_C} varies over the interval ${\left[C^\ga_{s}, C^\ga_0\right]}$, and therefore a linear stability analysis around the steady-state solution accounts only for the flow properties across the shock region, ignoring the rest of the Buckley-Leverett solution. This assumption was implicit in many earlier studies \citep{yortsos1986,yortsos1989,chikhliwala1988}. \citet{riaz2004} tested this assumption by perturbing the entire range of saturations in the complete Buckley-Leverett profile and comparing the dispersion relations (growth rate vs wavenumber) for the perturbed saturations. Since any perturbed saturation above or below the shock saturation resulted in smaller growth rates, the authors concluded that instability is indeed governed by the shock region---making the choice of the steady-state solution as the base state a reasonable one. 

Finally, the expression for the pressure derivative can be derived from Equation~\ref{eq:ut_nond}:
\begin{align}
    \frac{dP_\ell}{d\xi} = -\frac{1}{\lambda_t}\left( u_t - \lambda_\rho \right) - f_g \frac{dP_c}{dC^\ga} \frac{dC^\ga}{d\xi},
\label{eq:basestate_pl}
\end{align}
with $C^\ga$ the solution for Equation~\ref{eq:basestate_C}.

\subsection{The eigenvalue problem}
Based on Equations~\ref{eq:matbal_pl_nond} and~\ref{eq:ut_nond}, the two differential equations that define the two-dimensional, moving-frame problem in $(\xi,y)$ is:
\begin{align}
    \nabla \cdot \left( \lambda_t \nabla P_\ell + \lambda_\rho \mathbf{e}_g + \lambda_c \nabla C^\ga \right) &= 0, \label{eq:gradut_nond} \\
    \frac{\partial C^\ga}{\partial t} -v_s\frac{\p C^\ga}{\p \xi} - \nabla \cdot \left( \bar{\lambda}^\ga \nabla P_\ell + \bar{\lambda}_\rho^\ga \bol{e}_g \right) - \nabla \cdot \left(\bsym{\Lambda}_t^\ga \nabla C^\ga \right) &= 0 .\label{eq:matbal_pl_nond2} 
\end{align}

For the linear stability analysis of the system above, we ($i$) decompose the primary unknowns ($C^\ga,P_\ell$) in terms of their base-state values and small wave-like disturbances, and ($ii$) linearize the differential equations around the same base values. The primary variables ($C^\ga,P_\ell$) can be written as:
\begin{equation}
    \begin{aligned}
        C^\ga(\xi,y,t) &= \ovl{C}(\xi) + \delta C(\xi, y, t) =  \ovl{C}(\xi) + \hat{c}(\xi)\, e^{iny+\sig t}, \\
        P_\ell(\xi,y,t) &= \ovl{P}(\xi) + \delta P(\xi, y, t) = \ovl{P}(\xi) + \hat{p}(\xi)\, e^{iny+\sig t}.
    \end{aligned}
\label{eq:pertubed_var}
\end{equation}
where $\ovl{C}$ and $\ovl{P}$ are the \textit{base-state} values, $\hat{c}$ and $\hat{p}$ are the \textit{disturbed} values---with subscript $\ell$ and superscript $\ga$ suppressed for convenience---, $n$ is the perturbation wave-number and $\sig$ is the growth rate of the perturbation. We assume the disturbance eigenfunctions, $\hat{c}(\xi)$ and $\hat{p}(\xi)$, decay at infinity:

\begin{equation}
    \lim_{\xi\to\pm\infty} \hat{c} = 0\quad \text{and}\quad \lim_{\xi\to\pm\infty} \hat{p}= 0.
\end{equation}

The linearization of the flow functions around $\ovl{C}$ is done by expanding these functions in a Taylor series and ignoring the non-linear terms:
\begin{equation}
    f(C^\ga) \approx f(\ovl{C}) + f'(\ovl{C}) \delta C,
\end{equation}
where $f'$ indicates the derivative of $f$ with respect to $C^\ga$.

By substituting Definition~\ref{eq:pertubed_var} in Equations~\ref{eq:gradut_nond} and~\ref{eq:matbal_pl_nond2}, linearizing the flow functions and ignoring the quadratic terms in $\delta C$ and $\delta P$, and observing that the base-state solutions for Equations~\ref{eq:gradut_nond} and~\ref{eq:matbal_pl_nond2} imply
\begin{align}
    \dv{}{\xi} \left( \lambda_t\dv{\ovl{P}}{\xi} + \lambda_\rho + \lambda_c \dv{\ovl{C}}{\xi} \right) &= 0, \\
    \dv{\ovl{C}}{t} - v_s \dv{\ovl{C}}{\xi} - \dv{}{\xi} \left( \ovl{\lambda}\dv{\ovl{P}}{\xi} + \ovl{\lambda}_\rho + \Lambda_{t,xx} \dv{\ovl{C}}{\xi} \right) &= 0,
\end{align}
we arrive at the \textit{perturbation} equations:
\begin{align}
    \dv{}{\xi} \left( \lambda_t\dv{\hat{p}}{\xi} + \lambda'_t\dv{\ovl{P}}{\xi}\hat{c} + \lambda'_\rho\hat{c} + \lambda'_c\dv{\ovl{C}}{\xi}\hat{c} + \lambda_c\dv{\hat{c}}{\xi} \right) - n^2\lambda_t\,\hat{p} - n^2\lambda_c\,\hat{c}& = 0, \label{eq:eigen1a} \\
    v_s\dv{\hat{c}}{\xi} + \dv{}{\xi} \left( \ovl{\lambda}\dv{\hat{p}}{\xi} + \ovl{\lambda}'\dv{\ovl{P}}{\xi}\hat{c} + \ovl{\lambda}'_\rho\hat{c} + \Lambda_{t,xx}'\dv{\ovl{C}}{\xi}\hat{c} + \Lambda_{t,xx}\dv{\hat{c}}{\xi} \right)&\nonumber \\
    -n^2\ovl{\lambda}\,\hat{p} - n^2\Lambda_{t,yy}\,\hat{c}& = \sig\hat{c}, \label{eq:eigen1b}
\end{align}
where $\Lambda_{t,xx}$ and $\Lambda_{t,yy}$ are the diagonal elements of the dispersion tensor $\bsym{\Lambda}_t$. The derivatives of $\ovl{C}$ and $\ovl{P}$ are computed directly from Equations~\ref{eq:basestate_C} and~\ref{eq:basestate_pl}. Once again, the superscript $\ga$ was dropped for convenience. 

Equations~\ref{eq:eigen1a} and~\ref{eq:eigen1b} define the \textit{eigenvalue problem} in terms of eigenfunctions $\hat{c}$ and $\hat{p}$, and eigenvalue $\sig$. Expanding on the derivatives, we can rewrite the eigenvalue problem as:
\begin{align}
    A_2\ddv{\hat{p}}{\xi} + A_1\dv{\hat{p}}{\xi} + A_0 \hat{p} + B_2\ddv{\hat{c}}{\xi} + B_1\dv{\hat{c}}{\xi} + B_0 \hat{c} &= 0, \label{eq:eigen2a}\\
    E_2\ddv{\hat{p}}{\xi} + E_1\dv{\hat{p}}{\xi} + E_0 \hat{p} + F_2\ddv{\hat{c}}{\xi} + F_1\dv{\hat{c}}{\xi} + F_0 \hat{c} &= 0, \label{eq:eigen2b}
\end{align}
where
\begin{equation*}
    \begin{aligned}
        A_2 &= \lambda_t,\quad A_1 = \lambda'_t \dv{\ovl{C}}{\xi},\quad A_0 = -n^2 \lambda_t,\quad B_2 = \lambda_c, \\
        B_1 &= \lambda'_t \dv{\ovl{P}}{\xi} + \lambda'_\rho + 2\lambda'_c\dv{\ovl{C}}{\xi}, \\
        B_0 &= \lambda''_t\dv{\ovl{C}}{\xi}\dv{\ovl{P}}{\xi} + \lambda'_t\ddv{\ovl{P}}{\xi} + \lambda''_\rho\dv{\ovl{C}}{\xi} + \lambda''_c{\left(\dv{\ovl{C}}{\xi}\right)}^2 + \lambda'_c\ddv{\ovl{C}}{\xi} - n^2\lambda_c, \\
        E_2 &= \ovl{\lambda},\quad E_1 = \ovl{\lambda}' \dv{\ovl{C}}{\xi},\quad E_0 = -n^2 \ovl{\lambda},\quad F_2 = \Lambda_{t,xx}, \\
        F_1 &= \ovl{\lambda}' \dv{\ovl{P}}{\xi} + \ovl{\lambda}'_\rho + 2\Lambda'_{t,xx}\dv{\ovl{C}}{\xi} + v_s,\\
        F_0 &= \ovl{\lambda}''_t\dv{\ovl{C}}{\xi}\dv{\ovl{P}}{\xi} + \ovl{\lambda}'_t\ddv{\ovl{P}}{\xi} + \ovl{\lambda}''_\rho\dv{\ovl{C}}{\xi} + \Lambda''_{t,xx}{\left(\dv{\ovl{C}}{\xi}\right)}^2 + \Lambda'_{t,xx}\ddv{\ovl{C}}{\xi} - n^2\Lambda_{t,yy} - \sig.
    \end{aligned}
\end{equation*}

The linear stability analysis of the moving-frame problem, defined by Equations~\ref{eq:gradut_nond} and~\ref{eq:matbal_pl_nond2}, consists in solving numerically the eigenvalue problem (\ref{eq:eigen2a}-\ref{eq:eigen2b}) for $\sig$, for a given wavenumber $n$.

\subsection{Numerical solution of the eigenvalue problem}\label{sec:numerical_solution}
Numerically solving the eigenvalue problem requires careful handling of Equations~\ref{eq:eigen2a} and~\ref{eq:eigen2b} around the transition from the two-phase to the pure-liquid region (at $z^\ga=x^\ga_e$), where the flow functions and base-state derivatives ($d\ovl{C}/d\xi$ and $d\ovl{P}/d\xi$) are \textit{discontinuous}. 

Solving the base-state equation is straightforward: we set the transition point, where $C^\ga_e=C^\ga(x^\ga_e)$, at $\xi=0$, and we integrate Equation~\ref{eq:basestate_C} in its \textit{upstream} ($\xi\leq 0$) and \textit{downstream} ($\xi\geq 0$) domains separately---where flow is in the two-phase and pure-liquid regions, respectively.

The eigenvalue problem is a more complicated case. The derivatives of $\hat{c}$ and $\hat{p}$ are discontinuous at the transition point, and therefore using a conventional method such as the finite-difference method (as employed by \citet{riaz2004}) fails to solve our eigenvalue problem. To overcome this difficulty, we adapt the approach of \citet{chikhliwala1988}, which used the \textit{matched initial value problems} method. The workflow consists of:
\begin{enumerate}
    \item Find the far-field solutions of Equations~\ref{eq:eigen1a} and~\ref{eq:eigen1b} for their upstream and downstream domains (two-phase and pure-liquid regions, respectively), as functions of arbitrary constants to compute arbitrary initial values,
    \begin{equation*}
        \bol{m}_{up,iv} = \left\{\hat{p}(\xi_{up}), \hat{p}'(\xi_{up}), \hat{c}(\xi_{up}), \hat{c}'(\xi_{up}) \right\},\quad \xi_{up} < 0,
    \end{equation*}
    and
    \begin{equation*}
        \bol{m}_{dw,iv} = \left\{\hat{p}(\xi_{dw}), \hat{p}'(\xi_{dw}), \hat{c}(\xi_{dw}), \hat{c}'(\xi_{dw}) \right\},\quad \xi_{dw} > 0,
    \end{equation*}
    where the \textit{prime} here denotes the derivative with respect to coordinate $\xi$.
    \item Integrate the ordinary differential equations (ODE) for two sets of arbitrary initial values on the upstream domain ($\xi=[\xi_{up}, 0]$), and two sets on the downstream domain ($\xi=[0, \xi_{dw}]$), to obtain four linearly independent solutions at the transition point ${\xi=0}$:
    \begin{align*}
        \bol{m}_{1} &=\left\{\hat{p}_1(0^-), \hat{p}'_1(0^-), \hat{c}_1(0^-), \hat{c}'_1(0^-) \right\},\\
        \bol{m}_{2} &=\left\{\hat{p}_2(0^-), \hat{p}'_2(0^-), \hat{c}_2(0^-), \hat{c}'_2(0^-) \right\},\\
        \bol{m}_{3} &=\left\{\hat{p}_3(0^+), \hat{p}'_3(0^+), \hat{c}_3(0^+), \hat{c}'_3(0^+) \right\},\\
        \bol{m}_{4} &=\left\{\hat{p}_4(0^+), \hat{p}'_4(0^+), \hat{c}_4(0^+), \hat{c}'_4(0^+) \right\}.
    \end{align*}
    \item The linear combination of each upstream/downstream pair forms the \textit{general solution} for the upstream/downstream domains,
    \begin{align*}
        \bol{m}_{up} &= \al_1 \bol{m}_1 + \al_2 \bol{m}_2 = \left\{\hat{p}_{up}^-, \hat{p}_{up}^{\prime-}, \hat{c}_{up}^-, \hat{c}_{up}^{\prime-} \right\},\\
        \bol{m}_{dw} &= \al_3 \bol{m}_3 + \al_4 \bol{m}_4 = \left\{\hat{p}_{dw}^+, \hat{p}_{dw}^{\prime+}, \hat{c}_{dw}^+, \hat{c}_{dw}^{\prime+} \right\},
    \end{align*}
    where superscripts ``-'' and ``+'' indicate that the solution is computed at $0^-$ or $0^+$, and $\al_i$ are arbitrary constants.
    
    Continuity requires that $\hat{p}_{up}^-=\hat{p}_{dw}^+$ and $\hat{c}_{up}^-=\hat{c}_{dw}^+$, but such constraint does not apply for the derivatives of $\hat{p}$ and $\hat{c}$, since $\hat{p}_i$ and $\hat{c}_i$ are discontinuous at $\xi=0$. To solve the eigenvalue problem we need to derive the \textit{jump conditions} for the derivatives at $\xi=0$ (e.g. $\Delta\left(\hat{p}_i^{\prime}\right)=\lim_{\epsilon\to0}{\left[\hat{p}_i^\prime\right]}_{-\epsilon}^{+\epsilon},\ \epsilon>0$). More to the point, we seek a pair of expressions that couples the derivatives of $\hat{p}$ and $\hat{c}$ to the left and right of $\xi=0$:
    \begin{equation*}
        \hat{p}_i^{\prime+} = f\left( \hat{p}_i^-, \hat{p}_i^{\prime-}, \hat{c}_i^-, \hat{c}_i^{\prime-},\ovl{C}(0) \right),
    \end{equation*}
    and
    \begin{equation*}
        \hat{c}_i^{\prime+} = g\left( \hat{p}_i^-, \hat{p}_i^{\prime-}, \hat{c}_i^-, \hat{c}_i^{\prime-},\ovl{C}(0) \right).
    \end{equation*}
    
    With that, we can match the general solutions at the transition point $\xi=0$ by defining:
    \begin{equation*}
        \bol{m}_1^* = \left\{\hat{p}_1^-, \hat{p}_1^{\prime+}, \hat{c}_1^-, \hat{c}_1^{\prime+} \right\},
    \end{equation*}
    and
    \begin{equation*}
        \bol{m}_2^* = \left\{\hat{p}_2^-, \hat{p}_2^{\prime+}, \hat{c}_2^-, \hat{c}_2^{\prime+} \right\},
    \end{equation*}
    and then write:
    \begin{equation*}
        \al_1 \bol{m}_1^* + \al_2 \bol{m}_2^* = \al_3 \bol{m}_3 + \al_4 \bol{m}_4.
    \end{equation*}
    
    Finally, the non-trivial solution is:
    \begin{equation}
        \begin{vmatrix}
            \hat{p}_1^- & \hat{p}_2^- & \hat{p}_3^+ & \hat{p}_4^+ \\
            \hat{p}_1^{\prime+} & \hat{p}_2^{\prime+} & \hat{p}_3^{\prime+} & \hat{p}_4^{\prime+} \\
            \hat{c}_1^- & \hat{c}_2^- & \hat{c}_3^+ & \hat{c}_4^+ \\
            \hat{c}_1^{\prime+} & \hat{c}_2^{\prime+} & \hat{c}_3^{\prime+} & \hat{c}_4^{\prime+} 
        \end{vmatrix} = 0.
    \label{eq:nontrivial}
    \end{equation}
\end{enumerate}

\subsubsection{Far-field solutions}
We are interested in the far-field asymptotic solutions of the perturbation equations (\ref{eq:eigen1a} and~\ref{eq:eigen1b}) for the \textit{far-upstream} ($\xi<\xi_{up}$) and \textit{far-downstream} ($\xi>\xi_{dw}$) domains, where $d\ovl{C}/d\xi$ approaches zero, and the base-state variable $\ovl{C}$ and the flow functions have approximately constant values. The perturbation equations can then be expressed for the far-field domains as:

\begin{align}
    \lambda_t \left(\ddv{\hat{p}}{\xi} - n^2 \hat{p} \right) + \lambda_c\ddv{\hat{c}}{\xi} + \left( \lambda'_t\dv{\ovl{P}}{\xi} + \lambda'_\rho + \lambda'_c\dv{\ovl{C}}{\xi}\right) \dv{\hat{c}}{\xi} - n^2 \lambda_c \hat{c}& = 0, \label{eq:asymp1a}\\
    \ovl{\lambda} \left(\ddv{\hat{p}}{\xi} - n^2 \hat{p} \right) + \Lambda_{t,xx}\ddv{\hat{c}}{\xi} + \left( v_s + \ovl{\lambda}'\dv{\ovl{P}}{\xi} + \ovl{\lambda}'_\rho + \Lambda'_{t,xx}\dv{\ovl{C}}{\xi} \right) \dv{\hat{c}}{\xi}& \nonumber\\
    -\left(\sig + n^2 \Lambda_{t,yy} \right) \hat{c}& = 0. \label{eq:asymp1b}
\end{align}

By substituting Equation~\ref{eq:asymp1a} in~\ref{eq:asymp1b}, we obtain the homogeneous ODE for overall concentration $\hat{c}$:

\begin{equation}
    \quad\theta_2\ddv{\hat{c}}{\xi} + \theta_1\dv{\hat{c}}{\xi} + \theta_0\hat{c} = 0,
\end{equation}
where
\begin{equation*}
\def\m{\frac{\ovl{\lambda}}{\lambda_t}}
    \begin{aligned}
        &\theta_2 = \Lambda_{t,xx} - \m \lambda_c, \\
        &\theta_1 = v_s + \ovl{\lambda}'\dv{\ovl{P}}{\xi} + \ovl{\lambda}'_\rho + \Lambda'_{t,xx}\dv{\ovl{C}}{\xi} - \m \left( \lambda'_t\dv{\ovl{P}}{\xi} + \lambda'_\rho + \lambda'_c\dv{\ovl{C}}{\xi}\right), \\
        &\theta_0 = -\sig + n^2 \left( \m \lambda_c - \Lambda_{t,yy} \right).
    \end{aligned}
\end{equation*}

The boundary condition for the far-field domains are:
\begin{equation}
    \lim_{\xi \to -\infty} \hat{c}_{up}\left(\xi\right) = 0,\quad \text{and}\quad \lim_{\xi \to +\infty} \hat{c}_{dw}\left(\xi\right) = 0.
\end{equation}

The general solutions---up to an arbitrary constant---for $\hat{c}$ in the far-upstream and far-downstream domains are:
\begin{equation}
\begin{aligned}
    \hat{c}_{up} &= A_1 e^{r_{pos} \xi}, \\
    \hat{c}_{dw} &= A_2 e^{r_{neg} \xi},
\end{aligned}
\end{equation}
where $r_{pos}$ and $r_{neg}$ are the positive and negative roots of the characteristic equation,
\begin{equation}
     \theta_2 r^2 + \theta_1 r + \theta_0 = 0.
\end{equation}

By substituting the exponential solution for $\hat{c}$ in Equation~\ref{eq:asymp1a}, we obtain the non-homogeneous ODE for the pressure $\hat{p}$:
\begin{equation}
    \ddv{\hat{p}}{\xi} - n^2\dv{\hat{p}}{\xi} = -\frac{A}{\lambda_t}
    \left[
    \lambda_c r^2 + \left( \lambda'_t\dv{\ovl{P}}{\xi} + \lambda'_\rho + \lambda'_c\dv{\ovl{C}}{\xi}\right) r - n^2 \lambda_c 
    \right] e^{r\xi}.
\end{equation}

The boundary conditions for the far-field domains are:
\begin{equation}
    \lim_{\xi \to -\infty} \hat{p}_{up}\left(\xi\right) = 0,\quad \text{and}\quad \lim_{\xi \to +\infty} \hat{p}_{dw}\left(\xi\right) = 0,
\end{equation}
and the general solutions for $\hat{p}$ in the far-upstream and far-downstream domains are:
\begin{equation}
\begin{aligned}
    \hat{p}_{up} &= B_1 e^{n \xi} + \frac{A_1}{\lambda_t}\, \frac{\lambda_c r^2_{pos} + \left( \lambda'_t\dv{\ovl{P}}{\xi} + \lambda'_\rho + \lambda'_c\dv{\ovl{C}}{\xi}\right) r_{pos} - n^2 \lambda_c }{r^2_{pos} - n^2}\, e^{r_{pos}\xi}, \\
    \hat{p}_{dw} &= B_2 e^{-n \xi} + \frac{A_2}{\lambda_t}\, \frac{\lambda_c r^2_{neg} + \left( \lambda'_t\dv{\ovl{P}}{\xi} + \lambda'_\rho + \lambda'_c\dv{\ovl{C}}{\xi}\right) r_{neg} - n^2 \lambda_c }{r^2_{neg} - n^2}\, e^{r_{neg}\xi}.
\end{aligned}
\end{equation}

\subsubsection{Coupling the left and right-side derivatives}
The coupling expressions for the left and right-side derivatives (jump conditions) are obtained by integrating the perturbation equations across the transition point ($\xi=0$):
\begin{align}
    {\left[ \lambda_t\dv{\hat{p}}{\xi} + \lambda'_t\dv{\ovl{P}}{\xi}\hat{c} + \lambda'_\rho\hat{c} + \lambda'_c\dv{\ovl{C}}{\xi}\hat{c} + \lambda_c\dv{\hat{c}}{\xi} \right]}_{-\Delta\xi}^{+\Delta\xi}\nonumber\\
    -\int_{-\Delta\xi}^{+\Delta\xi} \left( n^2\lambda_t\,\hat{p} + n^2\lambda_c\,\hat{c} \right) d\xi &= 0, \label{eq:coupling1a}
\end{align}
\begin{align}
    {\left[ v_s\hat{c} + \ovl{\lambda}\dv{\hat{p}}{\xi} + \ovl{\lambda}'\dv{\ovl{P}}{\xi}\hat{c} + \ovl{\lambda}'_\rho\hat{c} + \Lambda_{t,xx}'\dv{\ovl{C}}{\xi}\hat{c} + \Lambda_{t,xx}\dv{\hat{c}}{\xi} \right]}_{-\Delta\xi}^{+\Delta\xi}\nonumber\\
    -\int_{-\Delta\xi}^{+\Delta\xi} \left[ n^2\ovl{\lambda}\,\hat{p} + \left(\sig + n^2\Lambda_{t,yy}\right)\hat{c} \right] d\xi &= 0. \label{eq:coupling1b}
\end{align}

Assuming that the flow functions and the base-state derivatives ($d\ovl{C}/d\xi$ and $d\ovl{P}/d\xi$) can be discontinuous but are \textit{bounded} in the neighbourhood of $\xi=0$, and that $\hat{c}$ is continuous, by taking the limit $\Delta\xi\to0$, we obtain:
\begin{align}
    {\left[ \lambda_t\dv{\hat{p}}{\xi} + \lambda'_t\dv{\ovl{P}}{\xi}\hat{c} + \lambda'_\rho\hat{c} + \lambda'_c\dv{\ovl{C}}{\xi}\hat{c} + \lambda_c\dv{\hat{c}}{\xi} \right]}_{0^-}^{0^+} &= 0, \label{eq:coupling2a}\\
    {\left[ \ovl{\lambda}\dv{\hat{p}}{\xi} + \ovl{\lambda}'\dv{\ovl{P}}{\xi}\hat{c} + \ovl{\lambda}'_\rho\hat{c} + \Lambda_{t,xx}'\dv{\ovl{C}}{\xi}\hat{c} + \Lambda_{t,xx}\dv{\hat{c}}{\xi} \right]}_{0^-}^{0^+} &= 0. \label{eq:coupling2b}
\end{align}

Using the fact that flow functions $\lambda_t$ and $\ovl{\lambda}$ are continuous at $\xi=0$, we can write:
\begin{align}
    \lambda_{t0} \left(\hdiff{\hat{p}}{+} - \hdiff{\hat{p}}{-}\right) + 
    \hat{c}_0 \beta_1 + \lambda_{c+}\hdiff{\hat{c}}{+} - \lambda_{c-}\hdiff{\hat{c}}{-} &= 0, \label{eq:coupling3a} \\
    \ovl{\lambda}_0 \left(\hdiff{\hat{p}}{+} - \hdiff{\hat{p}}{-}\right) + 
    \hat{c}_0 \beta_2 + \Lambda_{t,xx+}\hdiff{\hat{c}}{+} - \Lambda_{t,xx+}\hdiff{\hat{c}}{-} &= 0, \label{eq:coupling3b}
\end{align}
where
\begin{equation*}
\begin{aligned}
    \beta_1 &= \lambda'_{t+} \hdiff{\ovl{P}}{+} - \lambda'_{t-}\hdiff{\ovl{P}}{-} +
    \lambda'_{\rho+} - \lambda'_{\rho-} + 
    \lambda'_{c+}\hdiff{\ovl{C}}{+} - \lambda'_{c-}\hdiff{\ovl{C}}{-}, \\
    \beta_2 &= \ovl{\lambda}'_+ \hdiff{\ovl{P}}{+} - \ovl{\lambda}'_-\hdiff{\ovl{P}}{-} +
    \ovl{\lambda}'_{\rho+} - \ovl{\lambda}'_{\rho-} +
    \Lambda_{t,xx+}'\hdiff{\ovl{C}}{+} - \Lambda_{t,xx-}'\hdiff{\ovl{C}}{-},
\end{aligned}
\end{equation*}
where the \textit{minus} and \textit{plus} sign subscripts indicate that functions are evaluated at $\xi=0^-$ and $\xi=0^+$, respectively. For the flow functions and base state derivatives, that translates to being evaluated at $\ovl{C}(z^\ga=x^{\ga}_{e-})$ and $\ovl{C}(z^\ga=x^{\ga}_{e+})$, respectively.

By substituting Equation~\ref{eq:coupling3a} into~\ref{eq:coupling3b}, we can write the relationship between the left-side and right-side derivatives of $\hat{c}$:
\begin{equation}
    \hdiff{\hat{c}}{+} = \frac{\Lambda_{t,xx-} - \al \lambda_{c-}}{\Lambda_{t,xx+} - \al \lambda_{c+}} \hdiff{\hat{c}}{-} - \frac{\hat{c}_0}{\Lambda_{t,xx+} - \al \lambda_{c+}}
    \left( \beta_2 - \al \beta_1 \right), \label{eq:coupling4a}
\end{equation}
where $\al = \lambda_{t0}/\ovl{\lambda}_0$.

The relationship between the left-side and right-side derivatives of $\hat{p}$ are simply:
\begin{equation}
    \hdiff{\hat{p}}{+} = \hdiff{\hat{p}}{-} - \frac{1}{\lambda_{t0}} \left( 
    \hat{c}_0 \beta_1 + \lambda_{c+}\hdiff{\hat{c}}{+} - \lambda_{c-}\hdiff{\hat{c}}{-}
    \right). \label{eq:coupling4b}
\end{equation}

Equations~\ref{eq:coupling4a} and~\ref{eq:coupling4b} allows us to solve Equation~\ref{eq:nontrivial} for eigenvalue $\sig$.

\subsubsection{Capillary function choice}\label{sec:dPc}
Since the mechanical dispersion on the two-phase region is zero, we can write the dissipative function on the upstream domain as:
\begin{equation}
    D_{t,xx}^{up} = \lambda \Delta c^\ga \dv{P_c}{S_g}\dv{S_g}{C^\ga}=\frac{\lambda_g \lambda_\ell}{\lambda_t}\dv{P_c}{S_g},
\end{equation}
where we used the result that $dC^\ga/dS_g = \Delta c^\ga$ on the two-phase region.

Now, for a given $P'_c$ function, it is possible that
\begin{equation*}
    \lim_{C^\ga\to C^\ga_e} D_{t,xx} = \lim_{S_g\to 0} \frac{\lambda_g \lambda_\ell}{\lambda_t} \dv{P_c}{S_g} = 0,
\end{equation*}
and from Equation~\ref{eq:basestate_C},
\begin{equation*}
    \lim_{D_{t,xx}\to 0} \dv{\ovl{C}}{\xi},
\end{equation*}
becomes \textit{unbounded}.

To avoid this result, we take inspiration in previous work---e.g.\ \citep{chikhliwala1988} and \citep{riaz2004}---and choose the Leverett J-function in Equation~\ref{eq:Pc} to have its derivative in the form:
\begin{equation}
    \dv{J_c}{S_g} = \frac{A_{pc}}{k_{rg}\, k_{r\ell}},
\end{equation}
where $A_{pc}$ is a multiplying constant.

The dimensionless capillary force derivative can then be written as:
\begin{equation}
    \dv{P_c}{S_g} = \frac{1}{k_{rg}\, k_{r\ell}} = \frac{1}{M} \frac{1}{\lambda_g\, \lambda_\ell},
\label{eq:dPc}
\end{equation}
where is the viscosity ratio $M=\mu_{\ell e}/\mu_{ge}$ on the two-phase region.

This choice for $P'_c$ means that the dissipative function $d\ovl{C}/d\xi$ is bounded in the neighbourhood of $C^\ga_e$. As pointed out by \citet{riaz2004}, it also has a physically meaningful relationship with the fluid saturations in the two-phase region, and has the added benefit of allowing comparison with past investigations on linear stability of immiscible, two-phase systems.

\section{Results}\label{sec:results}
We refer to the two components in the binary fluid as component $a$ (the lighter component) and component $b$ (the heavier component). Phase equilibrium properties were calculated using the Peng-Robinson (PR78) equation of state, and viscosities were computed using the corresponding states (CS) model described in \citep{pedersen2006}.

Two simplifying assumptions were made: ($i$) the liquid phase density remains constant in the pure-liquid region (as discussed in Section~\ref{sec:phase_eq}), and ($ii$) that the liquid viscosity in the pure-liquid phase follows and exponential function given by:
\begin{equation}
    \mu_\ell = \mu^b e^{A_\mu \left(z^b - 1\right)},\quad \text{for}\quad z^b > x^b_e,
\label{eq:mu_exp}
\end{equation}
with
\begin{equation}
    A_\mu = \frac{\ln\mu^b - \ln\mu_{\ell e}}{1 - x^b_e},
\end{equation}
where $\mu^b$ and $\mu_{\ell e}$ are the viscosities of the pure component $b$ and the liquid phase at the two-phase region, respectively, calculated using the CS model.

When both gravity and mass transfer are present, the instability of our two-phase, two-component model is controlled by the dimensionless functions $k_{rg}$, $k_{r\ell}$ and $\mu_\ell$, and the dimensionless quantities $M=\mu_{\ell e}/\mu_{ge}$, $\Delta_\rho^*$, $D_{\ell,xx}^*$, $R_a=a_L/a_T$ and $z^\ga_i$. The high number of dimensionless variables obscure the physical interpretation of the system. Therefore, to facilitate our analysis, we define a \textit{dimensional} base case, and work with dimensional and non-dimensional variables as appropriate for each case. To avoid confusion, dimensionless variables are explicitly denoted with a superscript ``*'' in the following sections.

We note that, unlike the immiscible case studied by earlier authors \citep{yortsos1986, yortsos1989, chikhliwala1988, riaz2004}, for partially miscible flow the maximum growth rate ($\sig_{max}$) and cutoff wavenumber ($n_{cut}$) values do not scale linearly with the capillary number ($N_{ca}$). This is readily apparent when we observe---from Equations~\ref{eq:normalization} and~\ref{eq:lenghtscale}---that the $N_{ca}$ appears explicitly in the expression for the dimensionless mechanical dispersion tensor $\bol{D}_\ell^*$. In fact, as we show in Section~\ref{sec:dispersivities}, $\bol{D}_\ell^*$ has a very non-linear impact on $\sig_{max}$ and $n_{cut}$.

\FloatBarrier\subsection{Base case}\label{sec:basecase}
For our base case fluid we chose a CO$_2$-decane mixture. Table~\ref{tab:eosparam} lists the EOS parameters for the components. Flow conditions and the porous medium properties for the base case are listed in Table~\ref{tab:resprop}. For the purpose of evaluating the fluid properties, pressure and temperature were assumed to be constant at ${50\,bar}$ and ${50^\circ C}$, respectively (see Section~\ref{sec:phase_eq}).

\begin{figure}[!htb]
    \centering
    \includegraphics[width=1\linewidth]{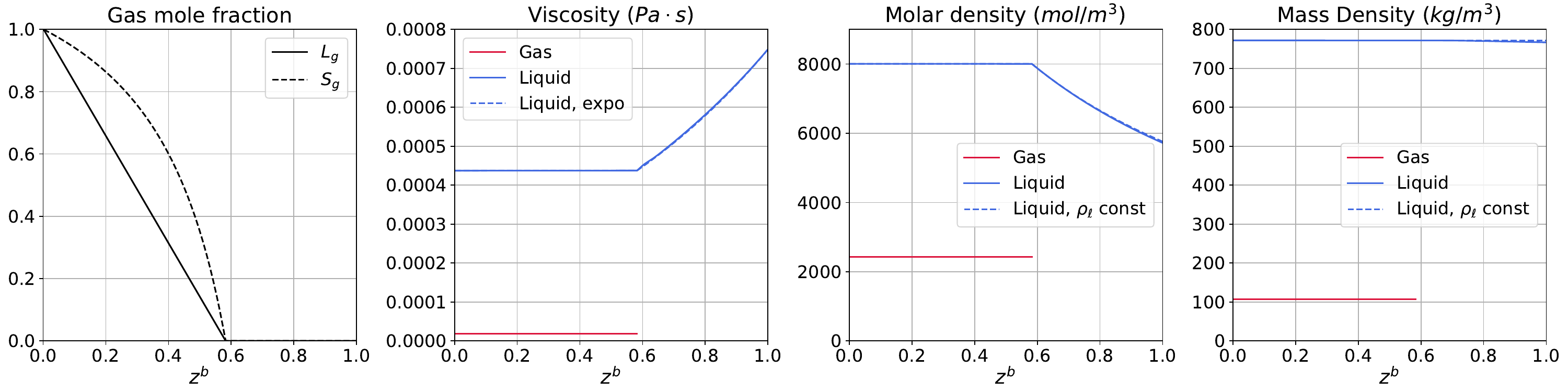}
    \caption{Fluid properties of the base case fluid at $P=50\, bar$ and $T=50^\circ C$.}\label{fig:fluidprop}
\end{figure}

Figure~\ref{fig:fluidprop} shows the gas and liquid properties for varying overall compositions (in terms of $z^b=1-z^a$), and a comparison between the viscosities, molar density and mass density with and without the simplifying assumptions above. It can be seen that, for our binary mixture, the assumption of constant liquid density in the pure-liquid region is reasonable, as the liquid density varies only slightly with composition, and that the exponential function for the liquid viscosity is a good approximation of the CS model. The viscosity ratio in the two-phase region is ${M_e=\mu_{\ell e}/\mu_{ge}\approx23}$, while the viscosity ratio between the initial liquid and injected gas is ${M_i=\mu_{\ell i}/\mu_{gi}\approx40}$.

\begin{table}[!htb]
    \centering
    \caption{EOS parameters for the components of the base case fluid.}\label{tab:eosparam}
    \begin{tabular}{l*{7}{c}}
        \toprule
        \multirow{2}{*}{\textbf{Component}} & $\bol{P_c}$ & $\bol{T_c}$ & $\textbf{Acc.}$ & $\bol{MW}$ & \multirow{2}{*}{$\bol{V_{shift}}$\footnotemark[1]} & \multicolumn{2}{c}{\textbf{BIC}} \\
        \cmidrule(r){7-8}
         & $(bar)$ & $(K)$ & \textbf{Factor} & $(g/mol)$ & & \textbf{CO}$\bol{_2}$ & \textbf{C10} \\
        \midrule
        CO$_2$ & 73.8 & 304.2  & 0.225 & 44  & -0.0817 & 0 & 0.115 \\
        C10 & 25.3 & 622.1  & 0.444  & 134 & 0.031  & 0.115  & 0\\
        \bottomrule
    \end{tabular}
    \footnotetext[1]{$V_{shift}$ is related to the volume translation correction term $c$ \citep{peneloux1982} by $c=bV_{shift}$, with $b$ the limiting volume parameter of the Peng-Robinson cubic equation.}
\end{table}

\begin{table}[!htb]
    \centering
    \caption{Reservoir properties and injection conditions at ${P=50\, bar,\ T=50^\circ C}$.}\label{tab:resprop}
    \begin{tabular}{*{8}{c}}
        \toprule
        $y^b_e$ & $x^b_e$ & $z^b_{inj}$ & $\bol{u_{inj}}$ & $\bol{\phi}$ & $\bol{k}$ & $\bol{a_L}$ & $\bol{a_T}$\\
        \midrule
        0.0013 & 0.5823 & 0.0013 & 0.1 $m^3/day$ & 0.1 & 100 mD & 10 mm & 2 mm \\
        \bottomrule
    \end{tabular}
\end{table}

\begin{figure}[!htb]
    \centering
    \includegraphics[width=0.8\linewidth]{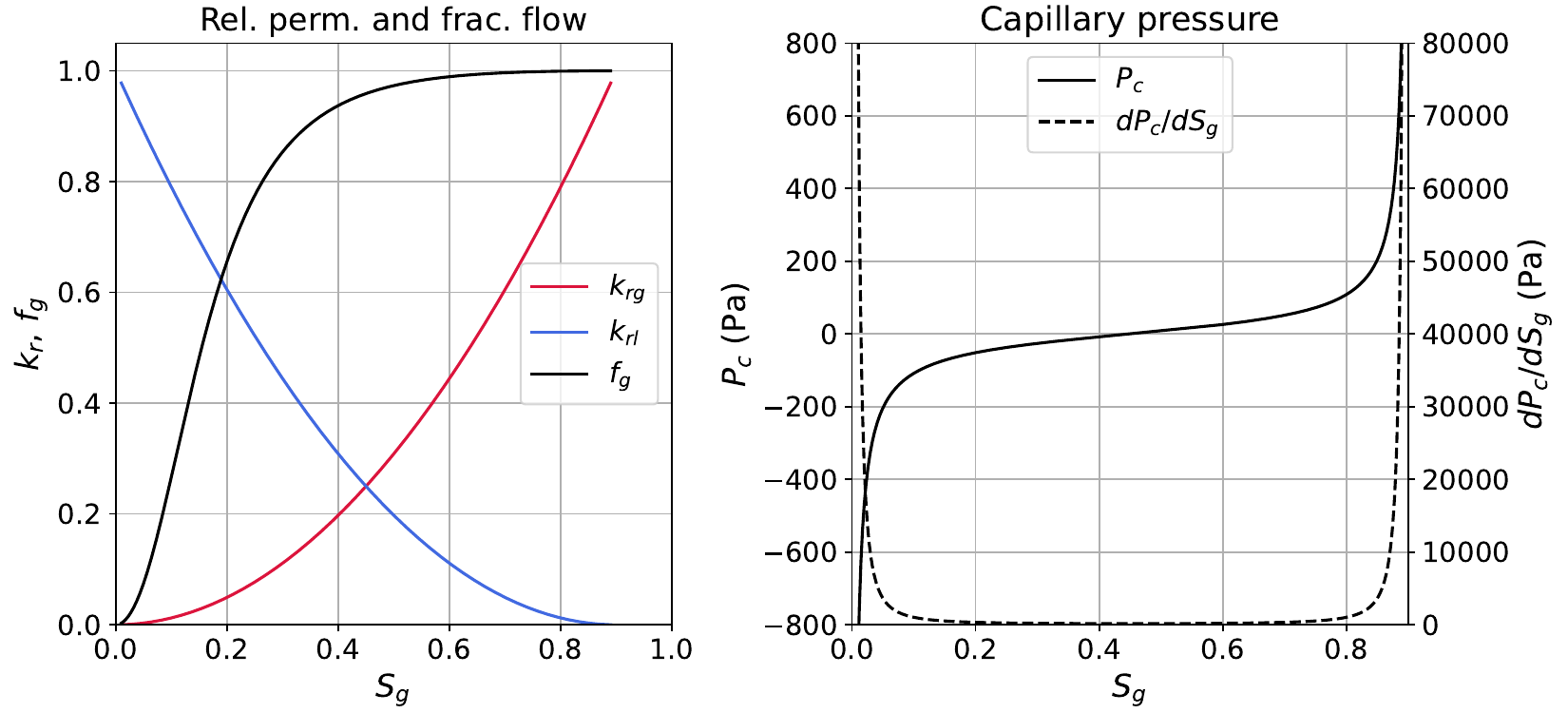}
    \caption{Relative permeabilities, fractional flow and capillary pressure for the base case.}\label{fig:rp_dpc1}
\end{figure}

The relative permeability (RP) parameters for the Corey-type base case RP are listed in Table~\ref{tab:rp1}. The capillary pressure is a function of the RPs, as described in Section~\ref{sec:dPc}, and the multiplying constant $A_{pc}$ is also listed in Table~\ref{tab:rp1}. Figure~\ref{fig:rp_dpc1} shows the RPs and the corresponding $P_c$ function.

\begin{table}[!htb]
    \centering
    \caption{Corey-type RP parameters and the $P_c$ multiplying constant for the base case.}\label{tab:rp1}
    \begin{tabular}{*{5}{c}}
        \toprule
        \textbf{Fluid} & $\bol{n}$ & $\bol{k_{rf}}$ & $\bol{S_r}$ & $\bol{A_{pc}}$ \\
        \midrule
        Gas & 2 & 1 & 0 & \multirow{2}{*}{0.001}\\
        Liquid & 2 & 1 & 0.1 \\
        \bottomrule
    \end{tabular}
\end{table}

In the following sections, most analyses consist of examining the sensitivity of stability to a set of chosen parameters. Unless otherwise stated, parameters not explicitly varied are held at their base case values.

\FloatBarrier\subsection{Initial composition}\label{sec:initialcomp}
The initial fluid composition ($z^b_i$) determines the amount of gas that can be dissolved into the liquid phase. If $z^b_i \leq x^b_e$, then the initial fluid is a saturated liquid and the displacement process occurs solely in the two-phase region---which, as shown in Section~\ref{sec:total_velocity}, means that the flow equations reduce to those of a two-phase, immiscible case. If $z^b_i > x^b_e$, then limited mass transfer is present in the displacement process, with flow being characterized as partially miscible, and transition from the two-phase to the pure-liquid region occurs at the shock. The lower the value of $x^b_i$, the greater the amount of mass transfer.

\begin{figure}[!htb]
    \centering
    \includegraphics[width=1\linewidth]{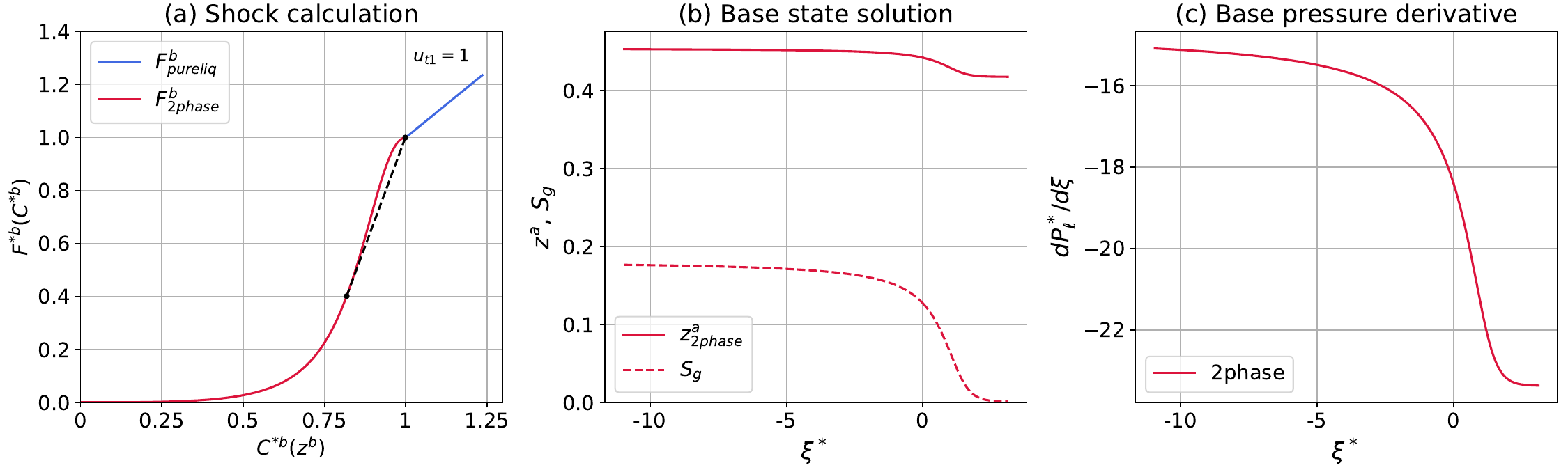}
    \caption{Saturated case ($z^b_i=0.582$ and $S_{gi}=0.001$): (a) Graphical representation of the shock calculation procedure; (b) the base-state solution, shown in terms of the overall composition of component $a$ and the gas saturation; (c) the base-state pressure derivative.}\label{fig:basestate_sgi001}
\end{figure}

\begin{figure}[!htb]
    \centering
    \includegraphics[width=1\linewidth]{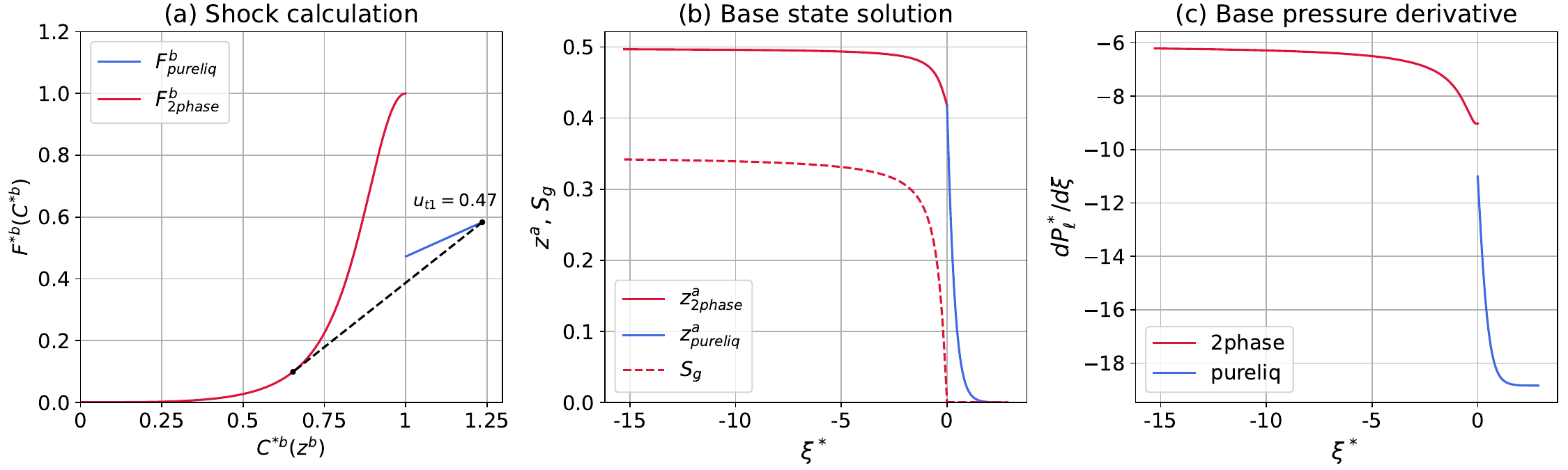}
    \caption{Same as Figure~\ref{fig:basestate_sgi001} for the partially miscible (PM) case ($z^b_i=1$).}\label{fig:basestate_zBi1}
\end{figure}

Figures~\ref{fig:basestate_sgi001} and~\ref{fig:basestate_zBi1} show the base-state profiles for two initial compositions: ($i$) $z^b_i=0.582$, just slightly below $x^b_e$, with $S_{gi}=0.001$, and ($ii$) $z^b_i=1$. For the former case, where flow is immiscible, the solution of the base-state equation (\ref{eq:basestate_C}) is smooth, while for the latter, where flow is partially miscible (PM), the base-state solution is not smooth and the base-state derivatives are discontinuous.

\begin{figure}[!htb]
    \centering
    \includegraphics[width=1\linewidth]{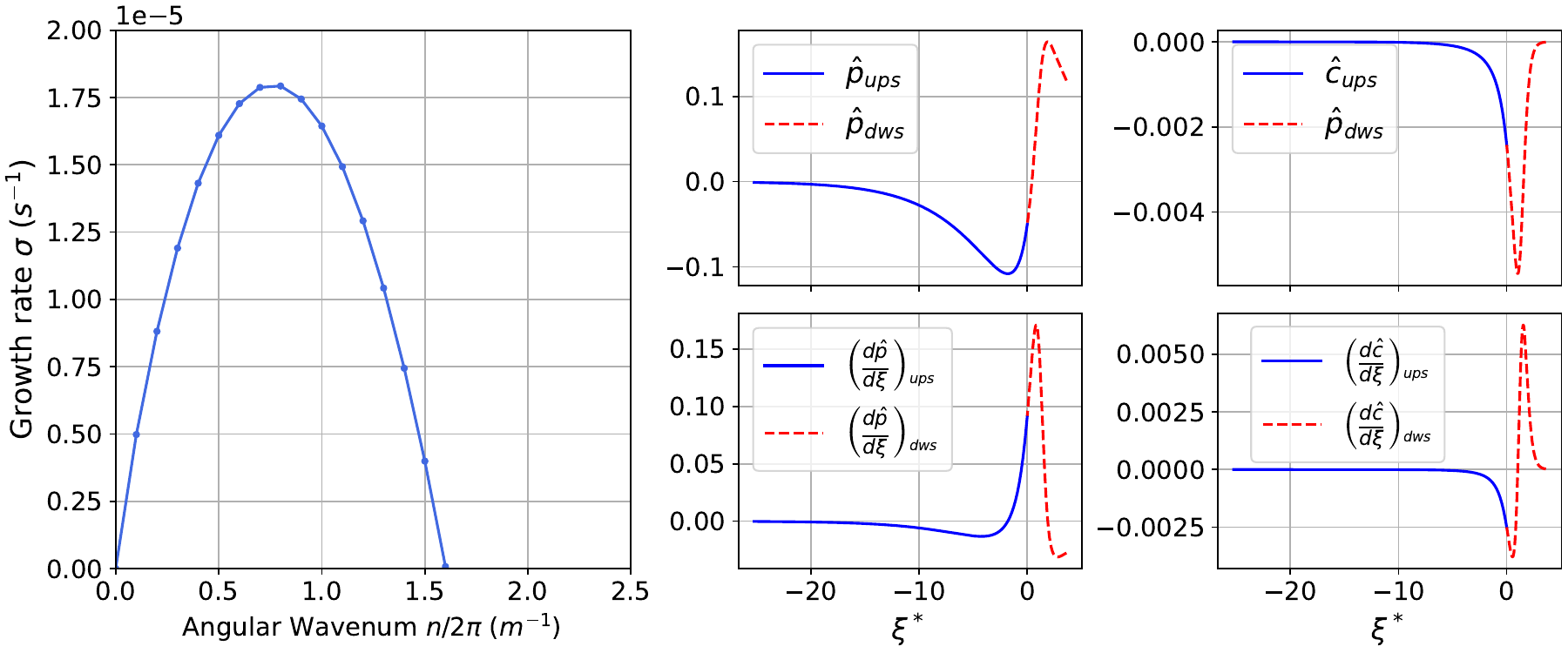}
\caption{Linear stability of the saturated case ($S_{gi}=0.001$). The leftmost graph shows the dispersion relation, and the remaining graphs shows eigenfunctions (in dimensionless form) for the most unstable mode ($n_{max}=0.76\, m^{-1}$).}\label{fig:insta_sgi001}
\end{figure}

\begin{figure}[!htb]
    \centering
    \includegraphics[width=1\linewidth]{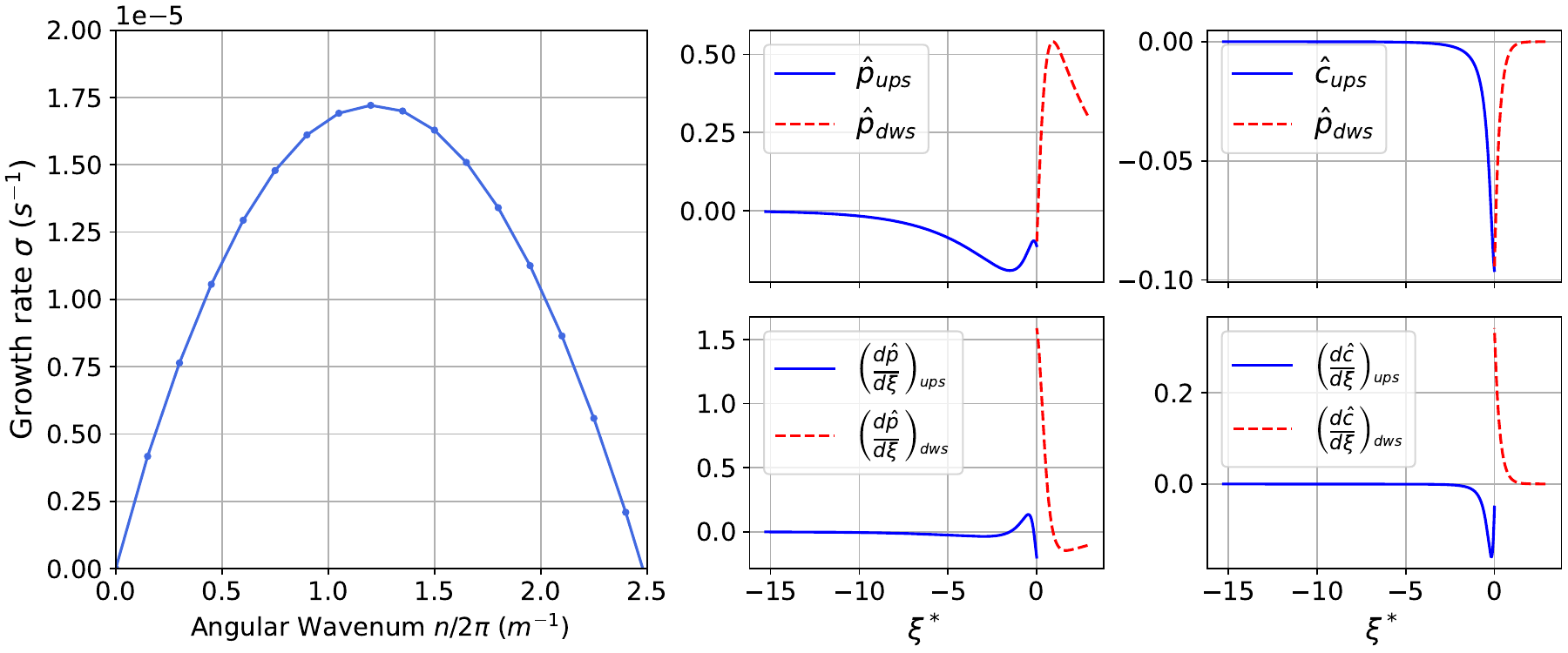}
\caption{Linear stability of the partially miscible (PM) case ($z^b_i=1$). Most unstable mode at (${n_{max}=1.2\, m^{-1}}$).}\label{fig:insta_zBi1}
\end{figure}

The results of the linear stability analysis for the two initial compositions are shown in Figures~\ref{fig:insta_sgi001} and~\ref{fig:insta_zBi1}, respectively, which show the plot of the growth rate vs wavenumber curves (dispersion relation) and the eigenfunctions associated with the maximum growth rate ($\sig_{max}$). We point out that while the eigenfunctions for the saturated case are continuous and smooth, those for the PM case are not (see Section~\ref{sec:numerical_solution}).

An interesting observation is that the maximum growth rate ($\sig_{max}$) for the PM case is lower than $\sig_{max}$ for the saturated case, while the opposite is true for the cutoff wavenumber ($n_{cut}$) and most unstable mode ($n_{max}$).

Many factors influence stability during the transition from saturated to undersaturated conditions (achieved by increasing $z^b_i$), making it difficult to isolate their individual effects. The most obvious impact is the increase in initial fluid viscosity and, consequently, in $M_i$. Another important effect is that increasing $z^b_i$ enhances mass transfer between phases, which in turn alters the shock configuration.

\begin{figure}
    \centering
    \includegraphics[width=1\linewidth]{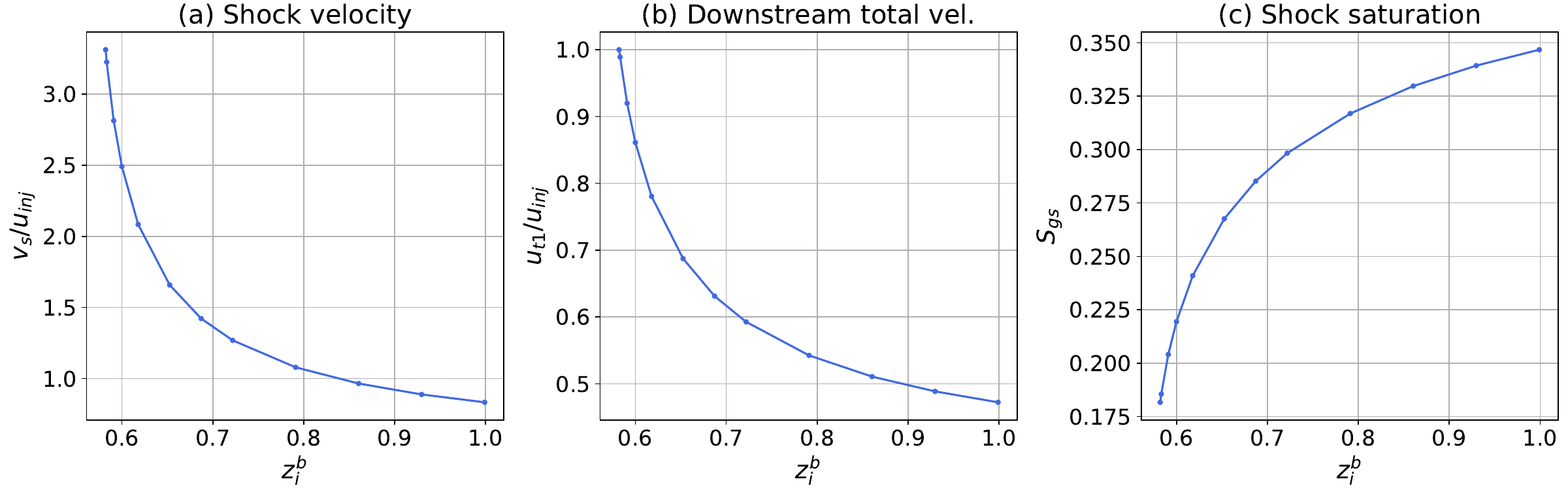}
    \caption{Shock velocity ($v_s$), downstream total velocity ($u_{t1}$) and shock gas saturation ($S_{gs}$) as a function of initial composition $z^b_i$. At $z^b_i\approx 0.582$, flow is immiscible and ${u_{t1}/u_{inj}=1}$.}\label{fig:zBi_shocks}
\end{figure}

To understand the impact of $z^b_i$ on the shock configuration, we first turn to the simplest case where the volume occupied by component $a$ remains unchanged as it dissolves in the liquid phase (i.e.\ $a$ has the same density as $b$), and the total velocities upstream ($u_{t0}$) and downstream ($u_{t1}$) of the shock are equal. From Figure~\ref{fig:basestate_sgi001}a, we see that an increase in the initial concentration of component $b$ ($C^b_i$) results in a decrease of $dF^b/dC^b$ at the shock, which in turn leads to a decrease of the overall concentration of $b$ at the shock ($C^b_s$). Because $C^b$ and the gas saturation are inversely related, an increase in $C^b_i$ means a higher saturation at the shock ($S_{gs}$), and finally, with the assumption that $u_t$ is constant, mass conservation requires that a higher $S_{gs}$ results in a lower shock front velocity $v_s$. In the more realistic case where component $a$ decreases in volume when transferring to the liquid phase, the total velocity downstream ($u_{t1}$) of the shock is lower than $u_{t0}$ (see Figure~\ref{fig:basestate_zBi1}a), and $S_{gs}$ is increased (and $v_s$ decreased) even further. These trends can be clearly observed in Figure~\ref{fig:zBi_shocks}.

The impact of these changes on stability can be inferred from the first-order approximation of the growth rate given by \citet{yortsos1986} for immiscible displacements:
\begin{equation}
    \frac{\sig}{n} = v_s \frac{\lambda_{t0} - \lambda_{t1}}{\lambda_{t0} + \lambda_{t1}}
\label{eq:1o_stab}
\end{equation}
where $\lambda_{t0}$ and $\lambda_{t1}$ are the total mobilities upstream and downstream of the shock, respectively. From Equation~\ref{eq:1o_stab}, we can expect a decrease in $v_s$ to reduce instability. Conversely, increasing $S_g$ at the shock should increase the mobility contrast ${\lambda_{t0}-\lambda_{t1}}$, which should increase instability. Additionally, the dispersion tensor $\bol{D}_\ell$ is linearly dependent on $u_{t1}$ and therefore $D_{\ell x}$ and $D_{\ell y}$ decrease with $z^b_i$. So we have competing effects, and the variation of $\bol{D}_\ell$ further complicates the picture.

\FloatBarrier\subsection{Dispersivities}\label{sec:dispersivities}
\begin{figure}[!htb]
    \centering
    \includegraphics[width=1\linewidth]{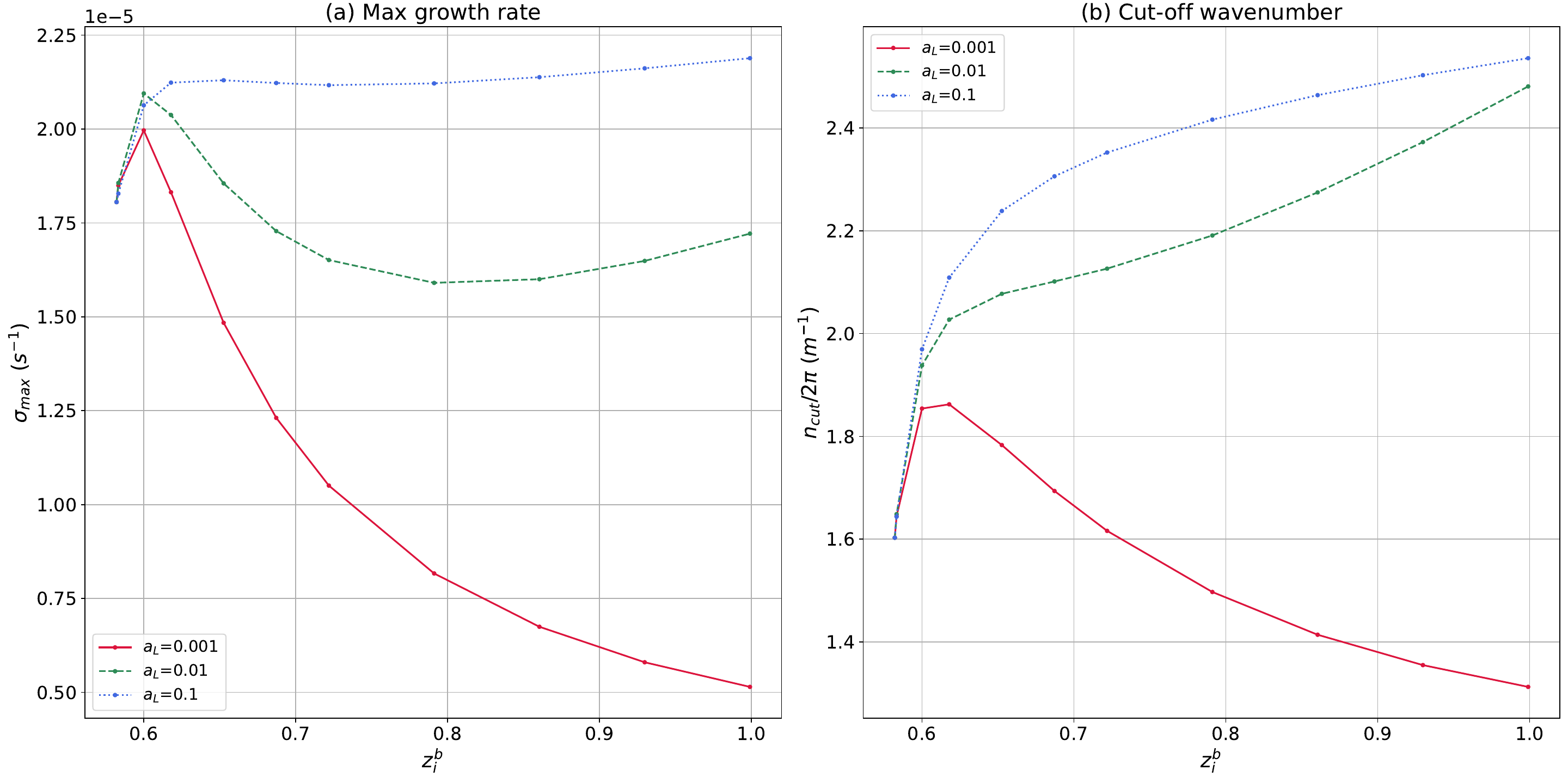}
    \caption{Influence of $z^b_i$ on $\sig_{max}$ and $n_{cut}$. Dispersivity values are given in meters. For all cases, $a_L/a_T=5$.}\label{fig:zBi_run_aR5}
\end{figure}

Figure~\ref{fig:zBi_run_aR5} shows the impact on stability of increasing mass transfer (by increasing $z^b_i$) for different values of for the longitudinal dispersivity $a_L$, with the ratio ${a_L/a_T=5}$. The dispersion relations are, in general, approximately symmetrical, and $n_{max}$ is approximately half of $n_{cut}$ for most cases. We see that $\sig_{{max}}$ and $n_{cut}$ exhibit a strongly non-linear behaviour. For the case where $a_L=0.001\,m$, both $\sig_{max}$ and $n_{cut}$ initially increase, and then decrease, with $z^b_i$. For the base case ($a_L=0.01\,m$), $\sig_{max}$ and $n_{cut}$ follow different trends, while the case $a_L=0.1\,m$ has the largest values for $\sig_{max}$ and $n_{cut}$ of all three cases, with $\sig_{max}$ varying only slightly with $z^b_i$. 

\begin{figure}
    \centering
    \includegraphics[width=1\linewidth]{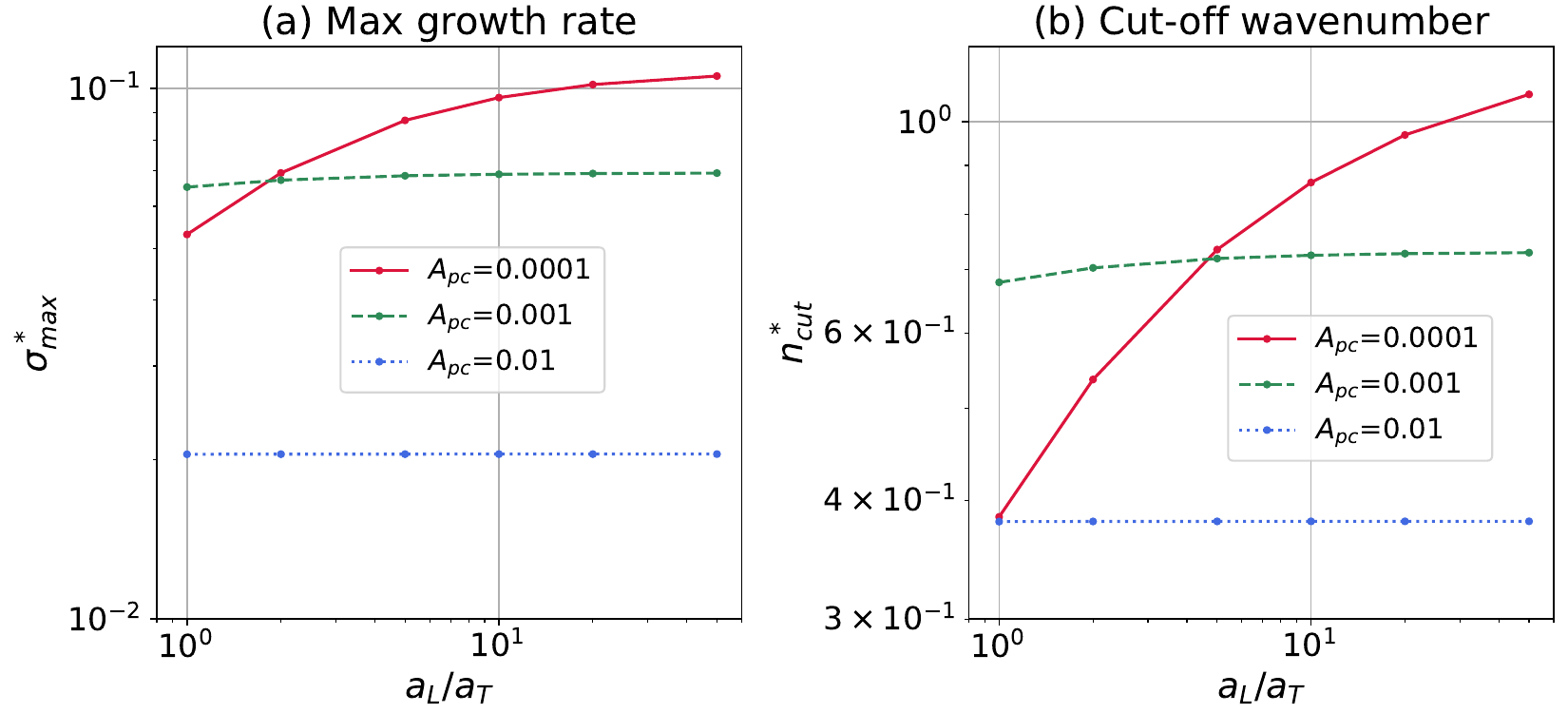}
    \caption{Impact of the transverse dispersivity $a_T$ on $\sig_{max}$ and $n_{cut}$. $a_L$ is fixed at $0.01\,m$, and $z^b_i=1$.}\label{fig:aT_run_aLfix}
\end{figure}

Intuitively, we would expect that higher values of $a_L$ would lead to a more stable system with lower values of $\sig_{max}$ and $n_{cut}$, but Figure~\ref{fig:zBi_run_aR5} shows the opposite for the cases selected. To understand the physics behind these results, we first need to understand the effect of transverse and longitudinal dispersivities on stability. To that end, we run a simple sensitivity on $a_L$ and $a_T$. First, transverse dissipative forces---be it from mechanical dispersion or capillary pressure---stabilize the displacement process by laterally dispersing the perturbations, and have a straightforward impact on stability. This is shown in Figure~\ref{fig:aT_run_aLfix}, where $a_T$ is varied while $a_L$ is held constant at $0.01\,m$. Three curves are shown for different values for the capillary pressure multipliers $A_{pc}$. As expected, increasing $a_T$ decreases both $\sig_{max}$ and $n_{cut}$, although the effect is quite limited for the cases where $A_{pc}=0.01\ \text{and}\ 0.001$, when mechanical dispersion is small compared to the capillary dispersion. 

\begin{figure}
    \centering
    \includegraphics[width=1\linewidth]{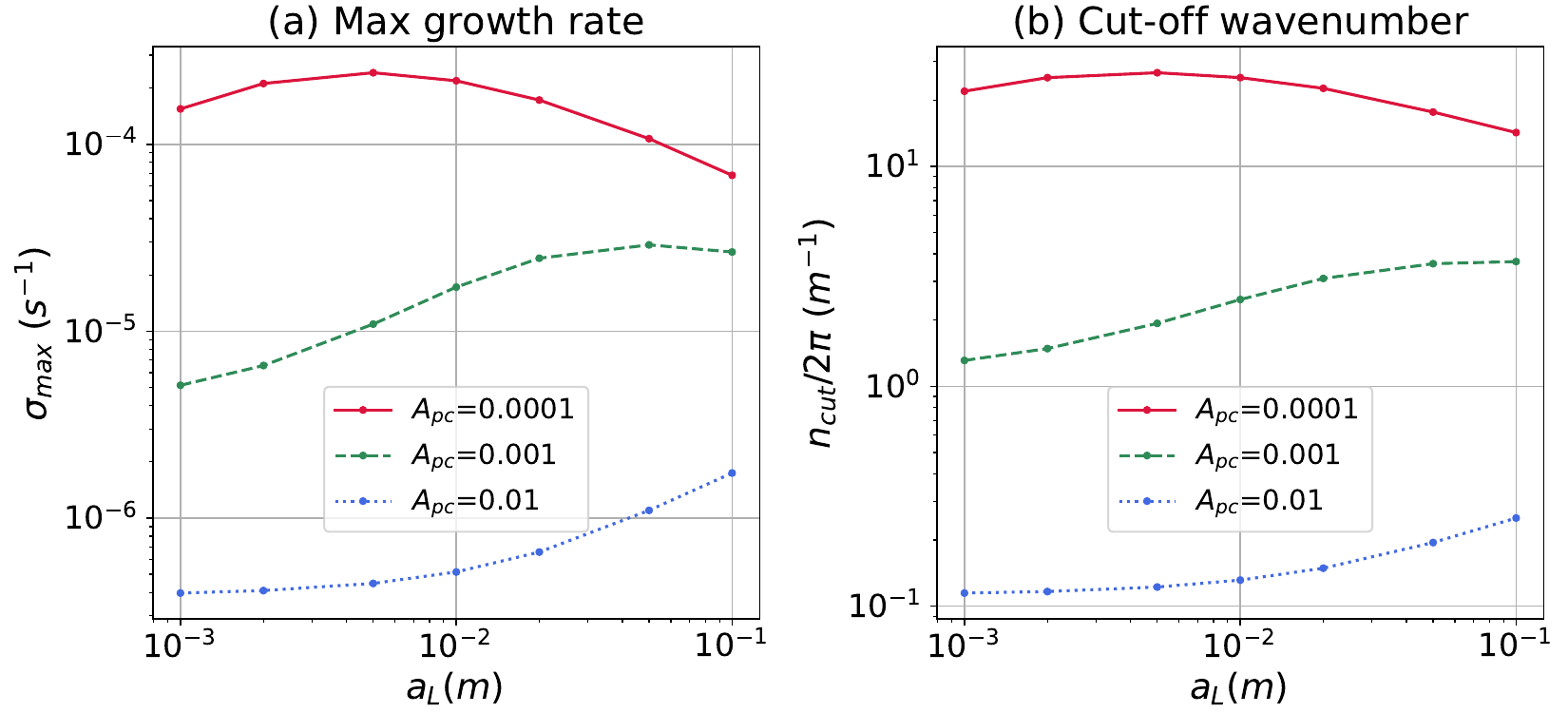}
    \caption{Impact of the longitudinal dispersivity $a_L$ on $\sig_{max}$ and $n_{cut}$. $a_T$ is fixed at $0.002\,m$, and $z^b_i=1$.}\label{fig:aL_run_aTfix}
\end{figure}

Longitudinal dissipative forces influence stability by smoothing the shock profile in the flow direction, reducing the overall concentration gradient of the base-state. Because it is the variation of the fluid properties along the displacement front that induces instability, one would expect that smoothing the $\ovl{C}$-profile would reduce instability. However, Figure~\ref{fig:aL_run_aTfix}, where the sensitivity for $a_L$ is plotted, shows a more complex picture. Two main observations emerge: ($i$) there appears to exist an intermediate value of $a_L$ for which instability is maximized, and ($ii$) $\sig_{max}$ and $n_{cut}$ exhibit asymptotic behaviour for low values of $a_L$ and large values of $P_c'$.

\begin{figure}
    \centering
    \includegraphics[width=1\linewidth]{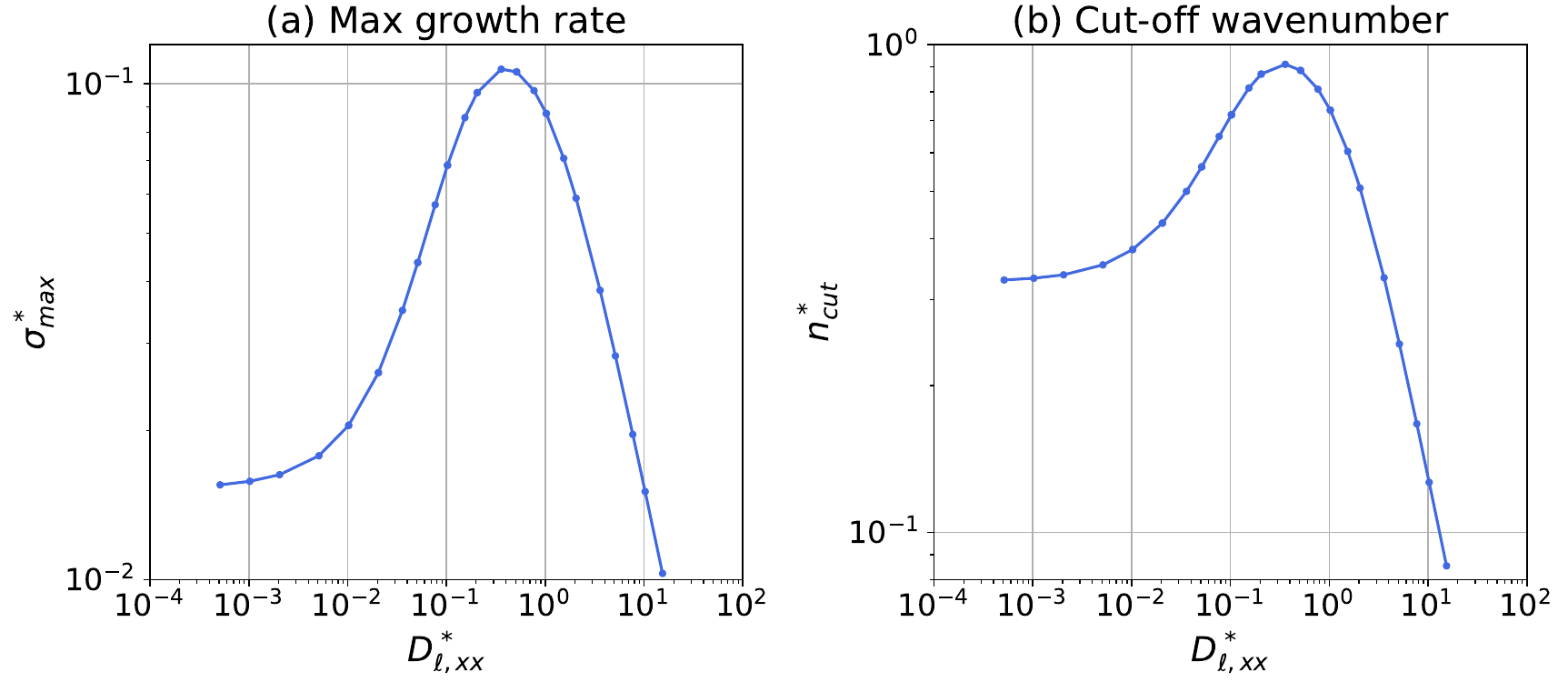}
    \caption{Plot of dimensionless $\sig_{max}^*$ and $n_{cut}^*$ versus $D^*_{\ell,xx}$. Initial composition is set at $z^b_i=1$, and $a_L/a_T=5$, for all cases.}\label{fig:Dstar_run}
\end{figure}

Observations ($i$) and ($ii$) are more easily seen in Figure~\ref{fig:Dstar_run}, where we plot the same simulations from Figure~\ref{fig:aL_run_aTfix} in \textit{dimensionless} form, with $\sig_{max}^*$ and $n_{cut}^*$ as functions of the dimensionless dispersion coefficient $D^*_{\ell,xx}$, recast in Equation~\ref{eq:Dstar} for convenience. Note that the expression for $D^*_{\ell,xx}$ incorporates the ratio between mechanical and capillary dispersion.

\begin{equation}
    D^*_{\ell,xx} = \frac{\mu_g}{\sqrt{\phi k}} \frac{a_L\, u_{t1}}{A_{pc}\, \sig_{g\ell}\,\cos\theta_{g\ell}}.
\label{eq:Dstar}
\end{equation}

Figure~\ref{fig:Dstar_run} shows that there exists an intermediate value of $D^*_{\ell,xx}$ for which instability is maximized ($D^{*\,max}_{\ell,xx}$). For lower values of $D^*_{\ell,xx}$, where mechanical dispersion is small compared to capillary dispersion, $\sig_{max}^*$ and $n_{cut}^*$ behave asymptotically. For higher values of $D^*_{\ell,xx}$, where mechanical dispersion dominates, $\sig_{max}^*$ and $n_{cut}^*$ fall steadily towards zero. This explains the behaviour of $\sig_{max}$ and $n_{cut}$ seen in Figure~\ref{fig:zBi_run_aR5}.

First, by calculating the values of $D^*_{\ell,xx}$ for the three Figure~\ref{fig:zBi_run_aR5} cases ({$a_L=0.001$, $0.01$ and $0.1\,m$}, with $D^*_{\ell,xx}\approx 10 a_T$ for the chosen parameters), we see from Figure~\ref{fig:Dstar_run} that, in agreement with Figure~\ref{fig:zBi_run_aR5}, the case $a_L=0.1\,m$ ($D^*_{\ell,xx}=1$) is indeed the most unstable case, and case $a_L=0.001\,m$ ($D^*_{\ell,xx}=0.01$) the most stable of the three.

Second, as discussed in Section~\ref{sec:initialcomp}, increasing $z^b_i$ has three effects on displacement: the shock velocity decreases, the mobility contrast increases (because $S_{gs}$ increases) and $D^*_{\ell,xx}$ decreases (because $u_{t1}$ decreases). As seen in Figure~\ref{fig:Dstar_run}, the impact on $\sig_{max}^*$ and $n_{cut}^*$ of decreasing $D^*_{\ell,xx}$ is strongly non-linear: if $D^*_{\ell,xx}<D^{*\,max}_{\ell,xx}$, then decreasing $D^*_{\ell,xx}$ should reduce instability, while if $D^*_{\ell,xx}>D^{*\,max}_{\ell,xx}$, decreasing $D^*_{\ell,xx}$ increases instability. This explains why, for the $a_L=0.001\,m$ ($D^*_{\ell,xx}=0.01$) case, $\sig_{max}^*$ and $n_{cut}^*$ mostly decrease with $z^b_i$, while the opposite happens for $a_L=0.1\,m$ ($D^*_{\ell,xx}=1$). Further discussion on the impact of $D^*_{\ell,xx}$ on stability is given in Section~\ref{sec:discussion}.

\FloatBarrier\subsection{Miscibility}\label{sec:miscibility}
Another variable that influences the extent of mass transfer between the two phases is the level of gas/liquid miscibility. A useful device to alter miscibility without changing the pressure, temperature or the components of the system is to alter the binary interaction coefficients (BIC) of the EOS model. Higher BIC values correspond to a lower content of component $a$ in the liquid phase at equilibrium, $x^a_e$---which we use as a proxy for system miscibility in this section.

\begin{figure}[!htb]
    \centering
    \includegraphics[width=1\linewidth]{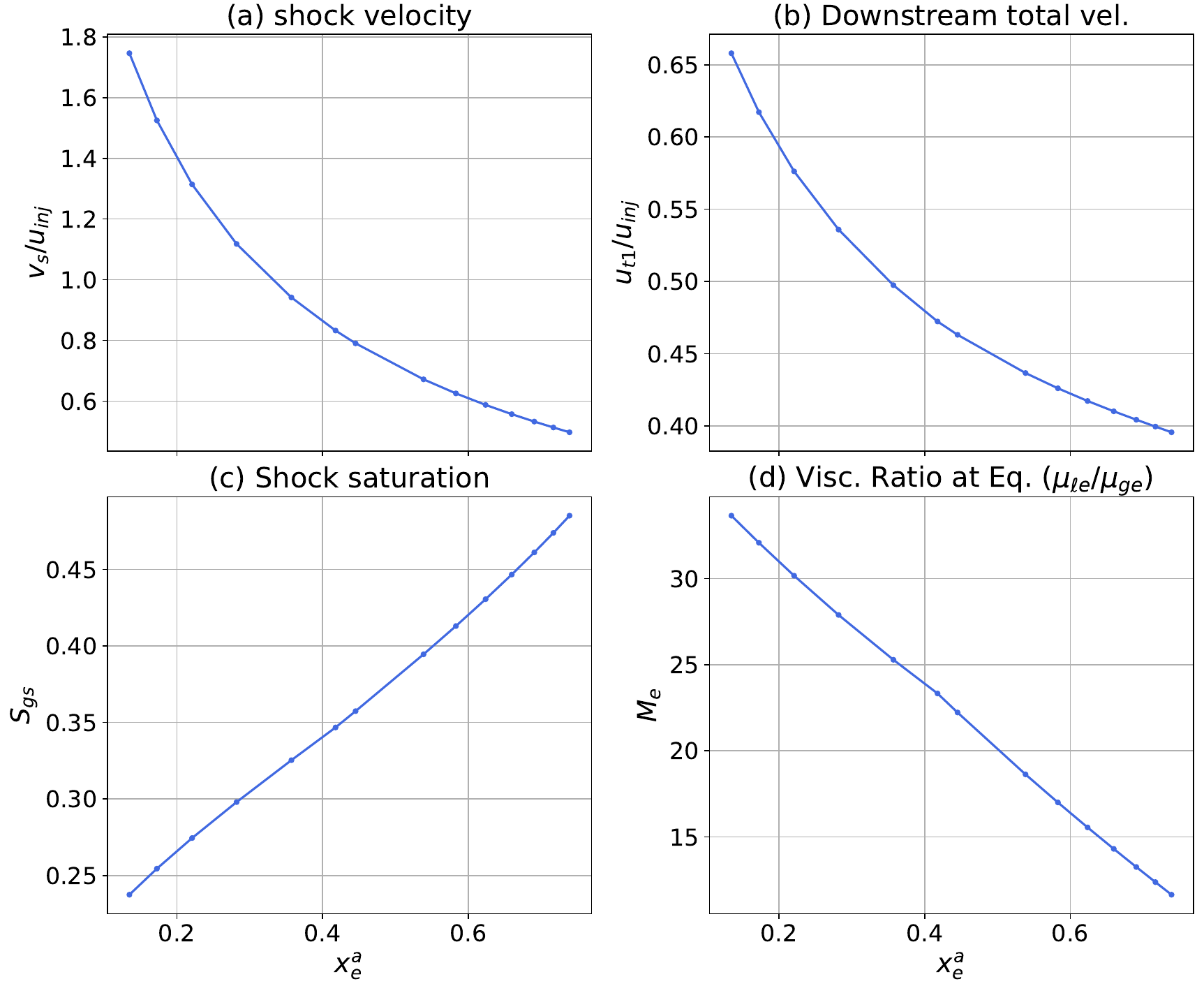}
    \caption{Shock velocity ($v_s$), downstream total velocity ($u_{t1}$), shock gas saturation ($S_{gs}$) and equilibrium viscosity ratio ($M_e=\mu_{\ell e}/\mu_{ge}$) as a function of the liquid overall composition in equilibrium ($x^a_e, 1-x^b_e$). For all cases, $z^b_i=1$ and $a_L/a_T=5$.}\label{fig:BIC_run_shocks}
\end{figure}

Figure~\ref{fig:BIC_run_shocks} shows the impact of increasing miscibility (or increasing $x^a_e$) on the shock configuration and the viscosity ratio of the two-phase region, $M_e$. It shows a similar picture to that of Figure~\ref{fig:zBi_shocks}, where we saw the changes in $v_s$, $u_{t1}$ and $S_{gs}$ due to increasing mass transfer. We note that the decrease in $M_e$ with $x^a_e$ further induce higher values of $S_{gs}$.

\begin{figure}[!htb]
    \centering
    \includegraphics[width=1\linewidth]{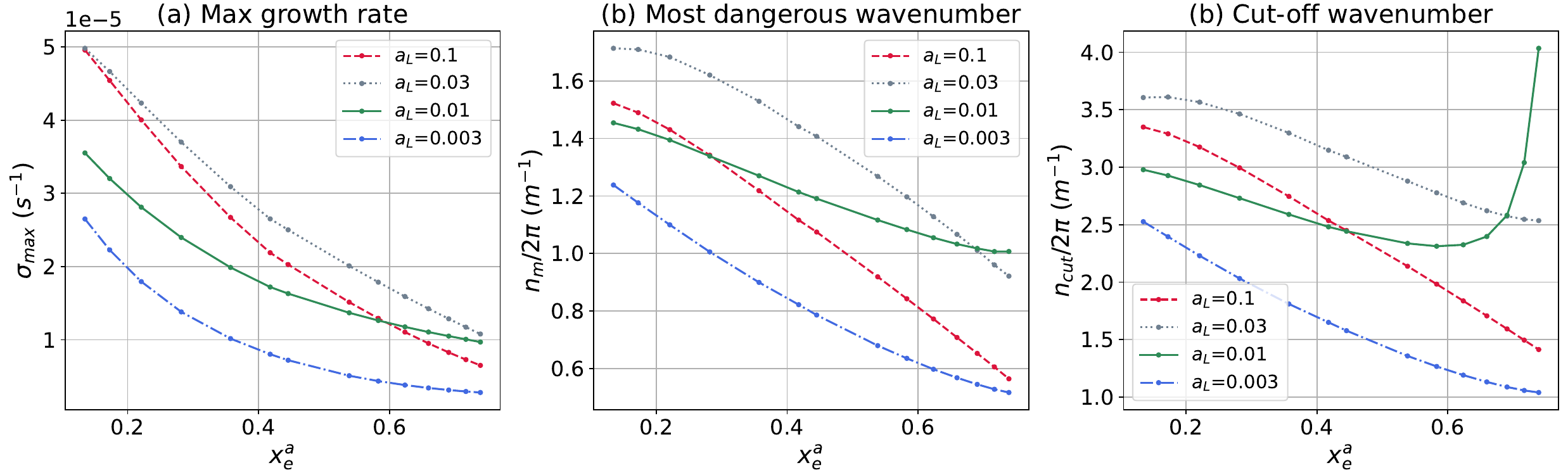}
    \caption{Plot of $\sig_{max}$ and $n_{cut}$ versus $x^a_e$. For all cases, $z^b_i=1$ and $a_L/a_T=5$.}\label{fig:BIC_run}
\end{figure}

As mentioned earlier, changes in the shock configuration have competing effects on stability, making it difficult to evaluate their impact. On the other hand, the decrease in $M_e$---which reduces the mobility contrast along the displacing front---has a clear stabilizing effect. Figure~\ref{fig:BIC_run} shows that, in all cases, the effect of increasing miscibility is to reduce $\sig_{max}$.

The behaviour of $n_{cut}$ is more complex. For the cases with $a_L=0.1$, 0.3, and 0.003, $n_{cut}$ also decreases with increasing miscibility. However, for the base case ($a_L=0.01$), $n_{cut}$ follows a very different trend at high values of $x^a_e$, increasing at an accelerating rate with $x^a_e$. We also observe that the approximate relation $n_{cut}\approx2n_{max}$ breaks down at high $x^a_e$, reflecting increased asymmetry in the dispersion relation curves. This divergent behaviour suggests a complex interplay between the various mechanisms governing system stability.

\FloatBarrier\subsection{Viscosity ratio}\label{sec:viscratio}
To analyse the impact of the viscosity ratio on stability, we run a series of cases where the viscosity of the liquid phase (in both the two-phase and pure-liquid regions) is altered by a multiplying constant, while keeping all other fluid properties of the base case unchanged. We compare two scenarios, one where immiscible displacement (IM) occurs, and one where flow is partially miscible (PM). The metric compared is the viscosity ratio between the initial and invading fluids, $M_i$.
 
\begin{figure}[!htb]
    \centering
    \includegraphics[width=1\linewidth]{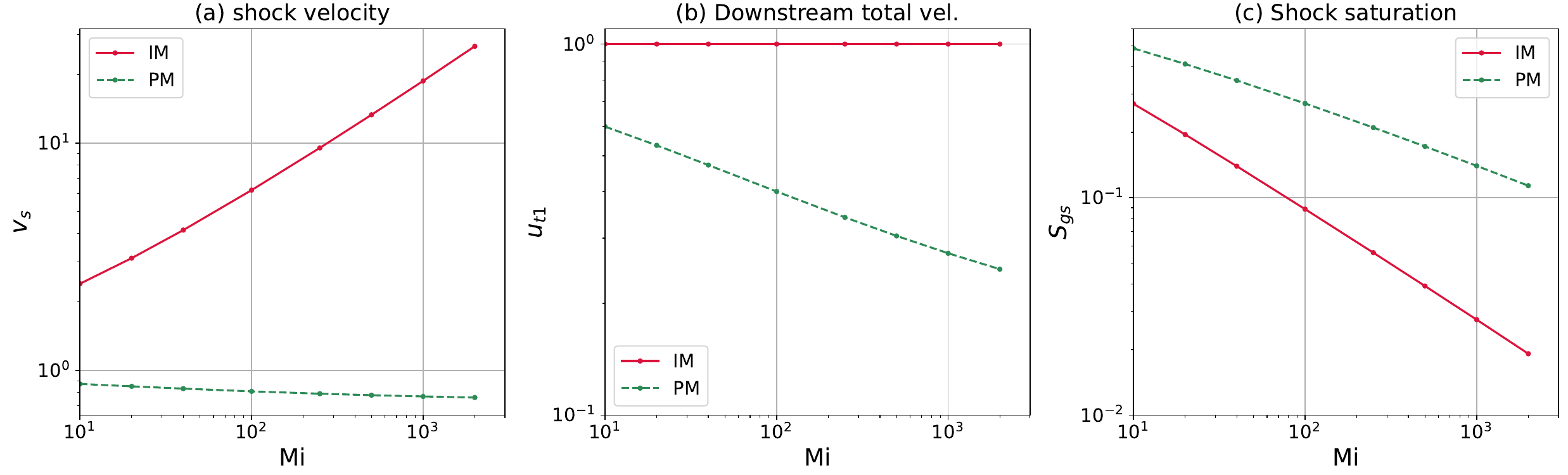}
    \caption{Shock velocity ($v_s$), downstream total velocity ($u_{t1}$) and shock gas saturation ($S_{gs}$) as a function of the viscosity ratio $M=\mu_\ell/\mu_g$. For all cases, $z^b_i=1$.}\label{fig:M_run_shocks}
\end{figure}

\begin{figure}[!htb]
    \centering
    \includegraphics[width=1\linewidth]{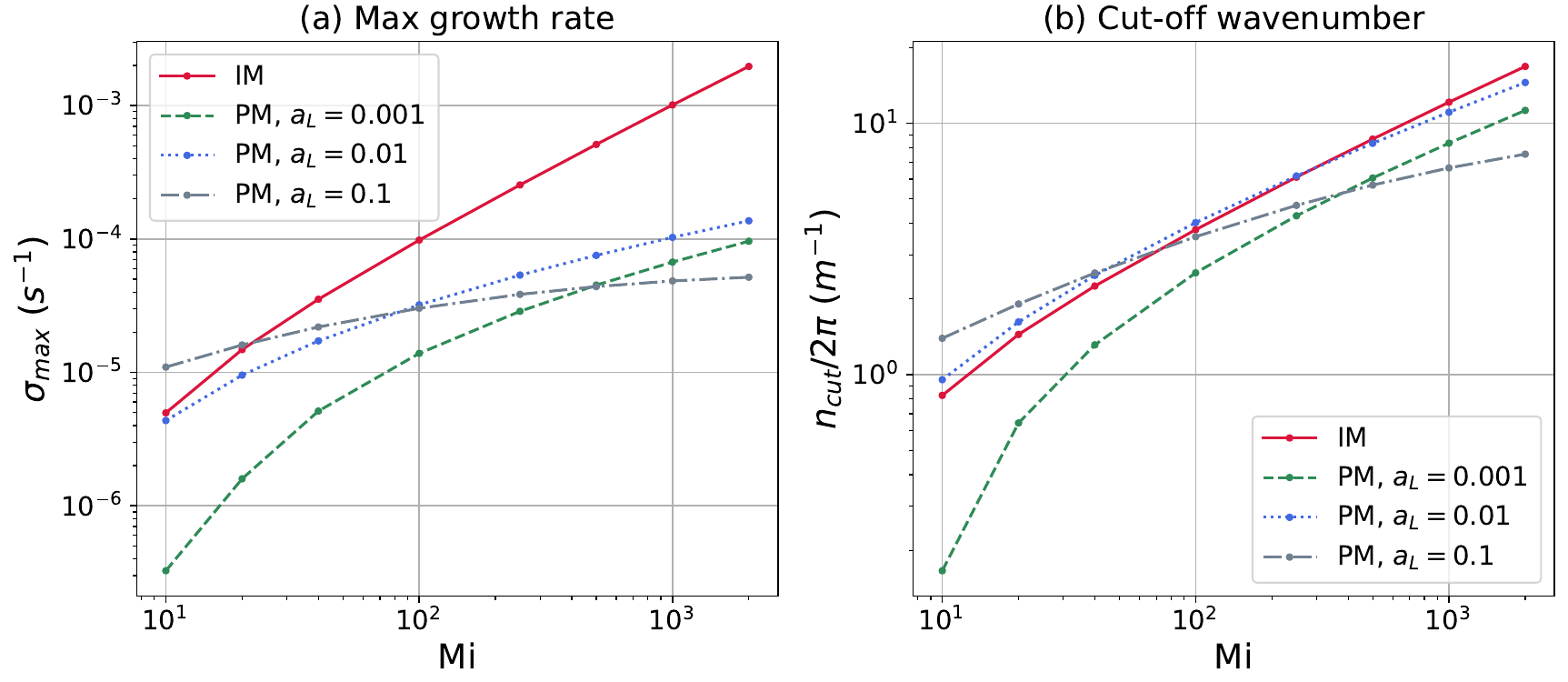}
    \caption{Plot of $\sig_{max}$ and $n_{cut}$ versus $M$. For all cases, $z^b_i=1$.}\label{fig:M_run}
\end{figure}

Figure~\ref{fig:M_run_shocks} compares the shock properties of the IM and PM cases, for various values~of~$M_i$, and in Figure~\ref{fig:M_run} we compare the instability of the IM and PM, for different values of $a_L$. The main observation in Figure~\ref{fig:M_run_shocks} is that, for the IM scenario, the shock velocity $v_s$ grows considerably with $M$, while in the PM scenario, $v_s$ fall slightly with $M$. This behaviour of $v_s$ explains why in the PM cases the slope of $\sig_{max}$ decreases with $M$, while for the IM case $\sig_{max}$ increases approximately linearly at high values of $M_i$ (as observed by \citet{riaz2004}). This \textit{dampening} effect is stronger for higher values of $a_L$, and we see that although the case with high $a_L$ ($0.001\,m$) is the most unstable at low $M_i$, at high values of $M_i$ it is the most stable. Finally, this dampening effect is less pronounced for the $n_{cut}$, and for the low $a_L$ case ($0.001\,m$) $n_{cut}$ increases with $M_i$ in a similar rate for both the IM and PM cases.

\FloatBarrier\subsection{Gravity}\label{sec:gravity}
Before we proceed with the analysis of gravity effects on stability, we look at the impact of the density variation alone (absent gravity, horizontal flow). The key here is that the density ratio between the gas and liquid phases in equilibrium ($\rho_{ge}/\rho_{\ell e}$) dictates the amount of change in volume that component $a$ undergoes when transferring from the gas to the liquid phase. The higher the density (mass or molar) of component $a$, the higher the density of the gas phase. When $c^a_e=c^b_e$, the phase densities match and component $a$ experiences no volume change upon transfer between the phases.

\begin{figure}[!htb]
    \centering
    \includegraphics[width=1\linewidth]{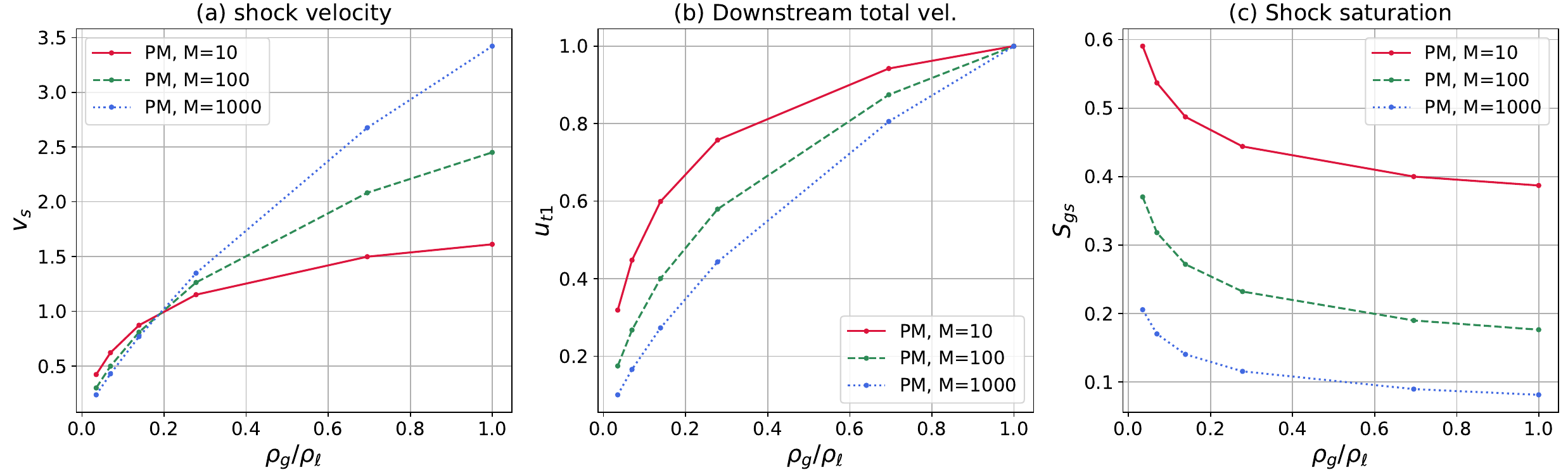}
    \caption{Shock velocity ($v_s$), downstream total velocity ($u_{t1}$) and shock gas saturation ($S_{gs}$) as a function of the gas/liquid density ratio, for horizontal flow. For all cases, $z^b_i=1$.}\label{fig:denV_run_shocks}
\end{figure}

As discussed earlier in Section~\ref{sec:initialcomp}, the lower the density of component $a$, the higher the shock saturation $S_{gs}$, and the lower the shock and downstream velocities, $v_s$ and $u_{t1}$. In Figure~\ref{fig:denV_run_shocks} we see these trends in all scenarios of $M_i$ (10, 100 and 1000), and can observe that $u_{t1}=1$ when $\rho_{ge}=\rho_{\ell e}$. 

\begin{figure}[!htb]
    \centering
    \includegraphics[width=1\linewidth]{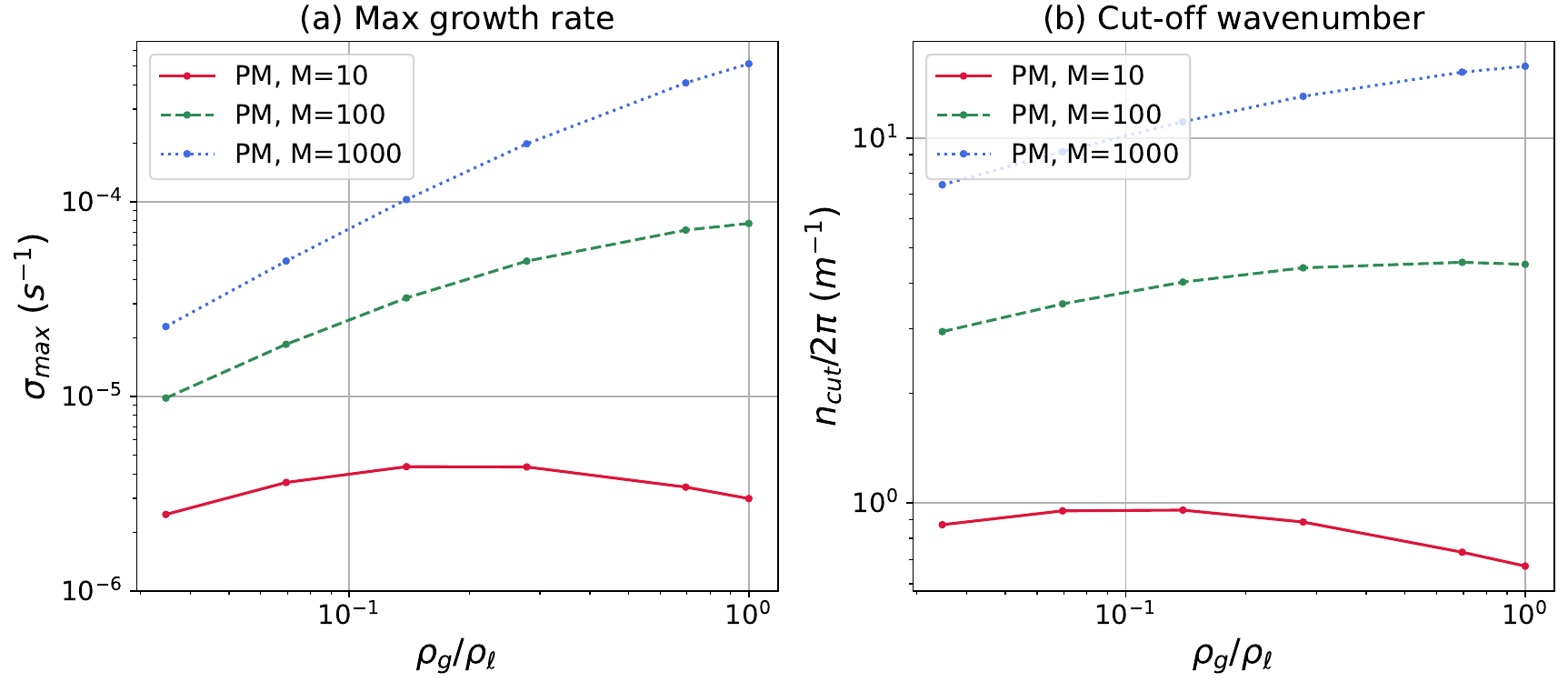}
    \caption{Plot of $\sig_{max}$ and $n_{cut}$ versus $\rho_{ge}/\rho_{\ell e}$. For the base case, $\rho_{ge}/\rho_{\ell e}=0.14$. For all cases, $z^b_i=1$.}\label{fig:denV_run}
\end{figure}

In Figure~\ref{fig:denV_run}, we see that the impact of volume change is higher for larger values of $M_i$, and that for cases $M_i=100$ and 1000, the effect of increasing $\rho_{ge}/\rho_{\ell e}$ is increasing $\sig_{max}$ and $n_{cut}$. For the case $M_i=10$, we find that the case with no volume change ($\rho_{ge}/\rho_{\ell e}=1$) is more stable than the base case ($\rho_{ge}/\rho_{\ell e}=0.14$). As discussed earlier, increasing $v_s$ increases instability, and decreasing $S_{gs}$ reduces the mobility contrast, decreasing instability. For the $M_i=10$ case, the latter appears to have a larger impact on overall instability.

Finally, we test the impact of gravity forces on stability. The two flow configurations studied are the vertically upward, where $\Delta\rho^*>0$, and the vertically downward ($\Delta\rho^*<0$) cases. In the case where $\Delta\rho^*=0$, both the upward and downward flow cases match the horizontal flow scenario. The variable $\Delta\rho^*$ was introduced in Equation~\ref{eq:phasevelo} and is recast in its dimensionless form in Equation~\ref{eq:Delta_rho} for clarity:
\begin{equation}
    \Delta\rho^* = \frac{k\, g}{u_{inj}\, \mu_{ge}} \left(\rho_{\ell e} - \rho_{ge}\right).
\label{eq:Delta_rho}
\end{equation}

\begin{figure}[!htb]
    \centering
    \includegraphics[width=1\linewidth]{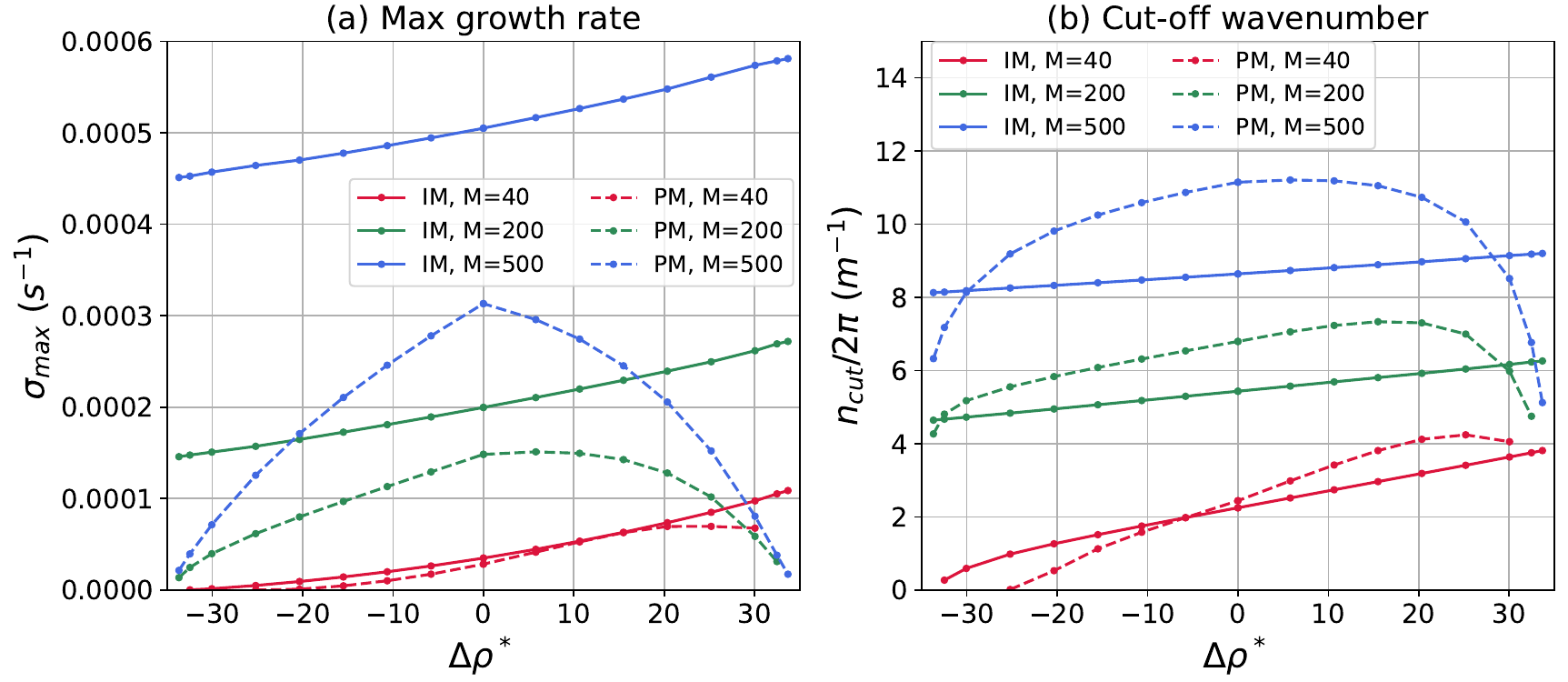}
    \caption{Plot of $\sig_{max}$ and $n_{cut}$ as a function of the dimensionless $\Delta\rho^*$. $\Delta\rho^*<0$ for downwards displacement, and $\Delta\rho^*>0$ for upwards displacement. For the immiscible (IM) case, $z^b_i=0.582$ and $S_{gi}=0.001$; for the partially miscible (PM), $z^b_i=1$.}\label{fig:UPDW_run}
\end{figure}

We analyse the impact of $\Delta\rho^*$ on stability for three different values of $M_i$ (40, 200 and 500), and for two types of displacement, IM and PM\@. The results are shown in Figure~\ref{fig:UPDW_run}. 

For the IM scenarios, viscous and gravity forces interact in a more additively fashion compared to the PM scenarios. For the former, we see that the curves $\sig_{max}$ vs $\Delta\rho^*$ for different viscosity ratios have similar monotonic shapes and are approximately parallel, whereas the PM curves are strongly non-linear functions of $\Delta\rho^*$. Interestingly, for the PM cases with high $M$ (100 and 1000), where viscous forces become more relevant than gravity, higher values of $|\Delta\rho^*|$ resulted in lower values of $\sig_{max}$ even for upward scenario ($\Delta\rho^*>0$). As discussed earlier, such behaviour is explained by mass exchange effects, with decreasing gas density decreasing instability (see Figure~\ref{fig:denV_run}). For the upward PM case with $M_i=40$, gravity does initially contribute to instability, with $\sig_{max}$ and $n_{cut}$ reaching maximum values at around $\Delta\rho^*\approx20$ and $\Delta\rho^*\approx25$, respectively. For the downward scenario, increasing ${|\Delta\rho^*|}$ always has a stabilizing influence, which is larger for the PM case (compared to the IM case), since $\Delta\rho^*<0$ not only introduces stabilizing gravity forces, but also mass-transfer effects.

\FloatBarrier\section{Discussion}\label{sec:discussion}
Comparison between immiscible and partially miscible displacements shows that mass transfer has a stabilizing effect on instability. In most cases presented in Section~\ref{sec:results}, adding or enhancing mass transfer reduces the maximum growth rate ($\sig_{max}$), particularly for displacements with large viscosity ratios (see Figure~\ref{fig:M_run}). Mass transfer also dampens gravity-induced instability and, in the analysed cases, exerts a larger stabilizing effect than gravity at large density and viscosity contrasts (see Figure~\ref{fig:UPDW_run}).

The impact on the cutoff wavenumber---$n_{cut}$, at which the displacement reaches marginal stability---is neither as pronounced nor as straightforward as that on $\sig_{max}$. For example, Figure~\ref{fig:M_run} shows that for the base case ($a_L=0.01$), $n_{cut}$ values for the IM and PM cases are quite similar. The most apparent effect of mass transfer is the alteration of the shock configuration---reducing the front velocity and increasing the invading fluid saturation---along with a reduction in the equilibrium viscosity ratio $M_e$. While the latter unequivocally promotes stability, the former introduces competing effects: reducing the shock velocity decreases the growth rate, but increasing the gas saturation at the shock increases mobility contrast, thereby promoting instability. One hypothesis is that while $v_s$ has less impact on $n_{cut}$ than on $\sig_{max}$, $S_{gs}$ exhibits the opposite behaviour, and if the decrease in $M_e$ caused by mass transfer is offset by the increase in $S_{gs}$, then both IM and PM scenarios would exhibit similar values for $n_{cut}$.

An interesting comparison can be made with results from \citet{tran2017}. Their study---which focused on marginal stability (i.e., $n_{cut}$) for a case with highly adverse displacement ($M_i\approx280$)---concluded that mass transfer promotes stability by reducing total flow velocity and viscosity contrast at the displacing front. While these mechanisms are consistent with those identified in the present study, we found that, although they significantly impact $\sig_{max}$, their effects on $n_{cut}$ are much less pronounced. In some cases, the PM scenario even exhibits larger $n_{cut}$ values than the IM scenario. A key difference in their model may explain this discrepancy: \citet{tran2017} assumed an idealized step-saturation displacement profile, thereby neglecting the effects of mass transfer on the shock profile.

Arguably, the most interesting result of our linear stability analysis for two-component compositional displacement is shown in Figure~\ref{fig:Dstar_run}, which illustrates the impact of $D^*_{\ell,xx}$ on instability. The behaviour at extreme values of $D^*_{\ell,xx}$ is physically plausible and intuitive: at very low values, capillary dispersion dominates and further reductions in $D^*_{\ell,xx}$ produce negligible changes in the displacement process, resulting in asymptotically small variations in $\sig_{max}$ and $n_{cut}$. At large values of $D^*_{\ell,xx}$, increasing dispersion stabilizes the displacement, as expected. However, the existence of a \textit{most dangerous} value $D^{*,max}_{\ell,xx}$---where the interaction between capillary and mechanical dispersion conspires to increase instability---is difficult to explain. To our knowledge, this complex behaviour has not been reported in the literature on viscous fingering and linear stability analysis.

The closest observation appears in \citet{jangir2021}, who compared models with constant and variable diffusion coefficients for nanoparticles. In the variable-diffusion case, the diffusion coefficient decreases rapidly with increasing concentration of other solutes, resulting in a \textit{sharper} base-state concentration profile compared to the constant-diffusion case (Figure 4 in \citep{jangir2021}). Interestingly, the dispersion relation showed that the variable-coefficient model---where concentration gradients are steeper---predicted lower values for both $\sig_{max}$ and $n_{cut}$. The authors attributed this finding to reduced maximum viscosity gradients in the transition zone under variable dispersion. Critically, the variable-coefficient model in their study affected only the concentration profile of stabilizing nanoparticles added to increase the invading fluid viscosity. In contrast, in our study the mobility contrast is either unaffected by $D^*_{\ell,xx}$ in the two-phase region or decreases monotonically with $D^*_{\ell,xx}$ in the single-phase region.

Four main caveats apply to the analysis presented in this paper. First, one main assumption in this analysis---and in several others in the literature---is that the relative permeability curves are identical for all values of $M$. This assumption is unlikely to hold in practice, as demonstrated by \citet{salmo2022}: although the viscous fingering experiments presented in their study showed a trend of decreasing shock saturations with increasing $M$, the authors had to select different sets of relative permeability curves for each experiment to capture these changes. Additionally, the linear scaling of $\sig_{max}$ with $M$ for large $M$, shown by \citet{riaz2004} (and in Figure~\ref{fig:M_run}), is actually valid only for a particular choice of relative permeability curves.

Second, our analysis in Section~\ref{sec:gravity} of gravitational effects on stability accounted for how the dimensionless density contrast---${\Delta\rho^*}$, given by Equation~\ref{eq:Delta_rho}---impacts the relative strength of viscous versus gravity forces and mass-transfer effects, which counteract gravity in upward displacement (see Figure~\ref{fig:UPDW_run}). However, a key assumption was that miscibility between the phases (or, equivalently, $x^b_e$) remains independent of density contrast, whereas in many practical cases these properties are inversely correlated---higher density contrasts typically correspond to lower miscibility. This assumption does not invalidate our analysis of how gravity and mass transfer affect instability, but it means Figure~\ref{fig:UPDW_run} does not fully represent real systems where mass transfer extent varies with density contrast.

Third, the assumption of constant liquid density in the pure-liquid region implies that gravity effects on instability arose from stabilizing/destabilizing forces in the two-phase region only. This simplifying assumption was reasonable for our base case given that $\rho_\ell$ varied little with $z^a_i$. However, this is not universally true, and it would be valuable to evaluate the impact of variable density in the liquid region. In fact, such a formulation may not add significant complexity to the two-component model. Although the continuity equation ${\nabla\cdot\bol{u}_t=0}$ would no longer be valid in the pure-liquid domain, the total mass balance equation---obtained by summing Equation~\ref{eq:matbal_pl} over $\ga$---could be used instead, since it is independent of Equation~\ref{eq:matbal_pl} and valid for non-constant densities.

Finally, mechanical dispersion is strongly scale-dependent, with laboratory experiments showing values of $a_L$ around $0.1$ to $10\,\text{mm}$, whereas field studies using numerical simulations show values up to $100\,\text{m}$ \citep{freeze1979}. Indeed, mechanical dispersion is fundamentally an upscaling technique \citep{moortgat2016}, and larger scales generally correspond to larger dispersivities. Since the dispersion relations may span vastly different scales depending on flow parameters, the appropriate dispersivity values may vary considerably. At larger scales and in weakly wet porous media, it may even be reasonable for mechanical dispersion to dominate over capillary effects.

\section{Conclusions}\label{sec:conclusion}
A mathematical model for two-phase, two-component displacement in porous media was successfully developed and implemented for linear stability analysis. The model accounts for partial miscibility, gravity effects, capillary forces, and mechanical dispersion in a system where a gaseous fluid displaces a heavier, more viscous liquid. By deriving the jump conditions for the eigenfunctions' derivatives---which are discontinuous at the transition from two-phase to pure-liquid flow at $\xi=0$---we were able to solve the eigenvalue problem numerically using the matched initial value problem (MIVP) method.

Results show that mass transfer in the partially miscible (PM) scenario reduces the equilibrium viscosity ratio ($M_e$), the shock front velocity ($v_s$), and the downstream total velocity ($u_{t1}$), while increasing the invading fluid saturation at the front ($S_{gs}$). The overall effect is stabilization of the displacement front, and increasing miscibility reduces both the maximum growth rate ($\sig_{max}$) and the cutoff wavenumber ($n_{cut}$) in most cases.

Comparison between immiscible (IM) and partially miscible (PM) displacement scenarios shows that mass transfer in PM flow dampens the effect of increasing viscosity ratio ($M$) on $\sig_{max}$, but has a much less pronounced impact on $n_{cut}$, particularly when dispersivities ($a_L$ and $a_T$) are small.

For upward displacement, the added instability from gravity forces is mitigated by mass-transfer effects, or even suppressed---the key insight being that the density contrast driving gravity-induced instability also enhances the extent of phase mass transfer. For downward displacement, both gravity and mass transfer act as stabilizing mechanisms.

An intriguing and unexpected finding is the existence of a \textit{most dangerous} value for the dimensionless longitudinal dispersion coefficient, $D^{*,max}_{\ell,xx}$, where both $\sig_{max}$ and $n_{cut}$ are maximized. This suggests a complex interaction between capillary forces, mechanical dispersion, and instability, and should be further investigated.

\backmatter%

\bmhead{Supplementary information} No supplementary information or material.

\bmhead{Acknowledgements} The authors thank Arne Skauge for his helpful remarks. The authors acknowledge the PhD scholarship for Paulo L. K. Caetano Chang provided by Petrobras. Kundan Kumar acknowledges funding from the Centre of Sustainable Subsurface Resources (CSSR), grant nr.\ 331841, supported by the Research Council of Norway, research partners NORCE Norwegian Research Centre and the University of Bergen, and user partners Equinor ASA, Harbour Energy Norge AS, Sumitomo Corporation, Earth Science Analytics, GCE Ocean Technology, and SLB Scandinavia.

\section*{Declarations}
\bmhead{Funding} Paulo Lee Kung Caetano Chang received funding from Petrobras.
Kundan Kumar received funding from Centre of Sustainable Subsurface Resources, Grant ID 331841.

\bmhead{Competing interests} The authors have no competing interests to declare that are relevant to the content of this article.

\bmhead{Ethics approval} Not applicable.

\bmhead{Consent for publication} Not applicable.

\bmhead{Data availability} Not applicable.

\bmhead{Materials availability} Not applicable.

\bmhead{Code availability} The python code developed to generate and solve the eigenvalue problem will be made available on request.

\bmhead{Author contribution} \textbf{P. L. K. Caetano Chang}: conceptualization; methodology; software; writing---original draft; writing---review \& editing. \textbf{K. Kumar}: conceptualization; methodology; writing---review \& editing.


\bibliography{references}

@book{bear2018,
  title={Modeling Phenomena of Flow and Transport in Porous Media},
  author={Jacob Bear},
  series={Theory and Applications of Transport in Porous Media},
  year={2018},
  publisher={Springer},
  address={Cham},
  doi={https://doi.org/10.1007/978-3-319-72826-1},
}

@book{orr2007,
author = {Orr, Franklin Mattes},
publisher = {Tie-Line},
address = {Holte},
isbn = {9788798996125},
year = {2007},
title = {Theory of gas injection processes},
language = {eng},
}

@article{saffman1958,
  title={The penetration of a fluid into a porous medium or Hele-Shaw cell containing a more viscous liquid},
  author={Saffman, Philip Geoffrey and Taylor, Geoffrey Ingram},
  journal={Proceedings of the Royal Society of London. Series A. Mathematical and Physical Sciences},
  volume={245},
  number={1242},
  pages={312--329},
  year={1958},
  publisher={The Royal Society London},
  doi={10.1098/rspa.1958.0085},
}

@article{hill1952,
  title={Channeling in packed columns},
  author={Hill, S },
  journal={Chemical Engineering Science},
  volume={1},
  number={6},
  pages={247--253},
  year={1952},
  publisher={Elsevier},
  doi={10.1016/0009-2509(52)87017-4},
}

@article{homsy1987,
  title={Viscous fingering in porous media},
  author={Homsy, George M},
  journal={Annual Review of Fluid Mechanics},
  volume={19},
  number={1},
  pages={271--311},
  year={1987},
  publisher={Annual Reviews 4139 El Camino Way, PO Box 10139, Palo Alto, CA 94303-0139, USA},
  doi={10.1146/annurev.fl.19.010187.001415},
}

@article{chuoke1959,
    author = {Chuoke, R.L. and van Meurs, P. and van der Poel, C.},
    title = "{The Instability of Slow, Immiscible, Viscous Liquid-Liquid Displacements in Permeable Media}",
    journal = {Transactions of the AIME},
    volume = {216},
    number = {01},
    pages = {188-194},
    year = {1959},
    month = {12},
    issn = {0081-1696},
    doi = {10.2118/1141-G},
    url = {https://doi.org/10.2118/1141-G},
}

@article{yortsos1986,
    author = {Yortsos, Y. C. and Huang, A. B.},
    title = "{Linear-Stability Analysis of Immiscible Displacement: Part 1---Simple Basic Flow Profiles}",
    journal = {SPE Reservoir Engineering},
    volume = {1},
    number = {04},
    pages = {378-390},
    year = {1986},
    month = {07},
    issn = {0885-9248},
    doi = {10.2118/12692-PA},
    url = {https://doi.org/10.2118/12692-PA},
}

@article{riaz2004,
    author = {Riaz, Amir and Tchelepi, Hamdi A.},
    title = "{Linear stability analysis of immiscible two-phase flow in porous media with capillary dispersion and density variation}",
    journal = {Physics of Fluids},
    volume = {16},
    number = {12},
    pages = {4727-4737},
    year = {2004},
    month = {12},
    issn = {1070-6631},
    doi = {10.1063/1.1812511},
    url = {https://doi.org/10.1063/1.1812511},
}

@article{peneloux1982,
author = {André Péneloux and Evelyne Rauzy and Richard Fréze},
title = {A consistent correction for Redlich-Kwong-Soave volumes},
journal = {Fluid Phase Equilibria},
volume = {8},
number = {1},
pages = {7-23},
year = {1982},
issn = {0378-3812},
doi = {https://doi.org/10.1016/0378-3812(82)80002-2},
url = {https://www.sciencedirect.com/science/article/pii/0378381282800022}
}

@Article{salmo2022,
  author   = {Salmo, I. C. and Sorbie, K. S. and Skauge, A. and Alzaabi, M. A.},
  journal  = {Transport in Porous Media},
  title    = {Immiscible Viscous Fingering: Modelling Unstable Water--Oil Displacement Experiments in Porous Media},
  year     = {2022},
  issn     = {1573-1634},
  month    = {Nov},
  number   = {2},
  pages    = {291-322},
  volume   = {145},
  day      = {01},
  doi      = {10.1007/s11242-022-01847-8},
  url      = {https://doi.org/10.1007/s11242-022-01847-8},
}

@article{moortgat2016,
  title={Viscous and gravitational fingering in multiphase compositional and compressible flow},
  author={Moortgat, Joachim},
  journal={Advances in Water Resources},
  volume={89},
  pages={53--66},
  year={2016},
  publisher={Elsevier},
  doi={10.1016/j.advwatres.2016.01.002}
}

@article{tran2017,
  title={Stability of {CO$_2$} displacement of an immiscible heavy oil in a reservoir},
  author={Tran, Truynh Quoc My Duy and Neogi, Parthasakha and Bai, Baojun},
  journal={SPE Journal},
  volume={22},
  number={2},
  pages={539--546},
  year={2017},
  publisher={Society of Petroleum Engineers},
  doi={10.2118/184407-PA}
}

@article{daripa2008,
  title={On capillary slowdown of viscous fingering in immiscible displacement in porous media},
  author={Daripa, Prabir and Pa{\c{s}}a, G.},
  journal={Transport in Porous Media},
  volume={75},
  number={1},
  pages={1--16},
  year={2008},
  publisher={Springer},
  doi={10.1007/s11242-008-9211-2}
}

@article{jangir2021,
  title={Linear stability analysis of miscible displacement by nanofluid with concentration-dependent diffusivity},
  author={Jangir, Pooja and Mohan, Ratan and Chokshi, Paresh},
  journal={Chemical Engineering Science},
  volume={240},
  pages={116609},
  year={2021},
  publisher={Elsevier},
  doi={10.1016/j.ces.2021.116609}
}

@article{tan1986,
  title={Stability of miscible displacements in porous media: Rectilinear flow},
  author={Tan, C. T. and Homsy, G. M.},
  journal={Physics of Fluids},
  volume={29},
  number={11},
  pages={3549--3556},
  year={1986},
  publisher={American Institute of Physics},
  doi={10.1063/1.865832}
}

@article{hickernell1986,
author = {Hickernell, F. J. and Yortsos, Y. C.},
title = {Linear Stability of Miscible Displacement Processes in Porous Media in the Absence of Dispersion},
journal = {Studies in Applied Mathematics},
volume = {74},
number = {2},
pages = {93-115},
doi = {https://doi.org/10.1002/sapm198674293},
year = {1986}
}

@article{hagoort1974,
    author = {Hagoort, J.},
    title = {Displacement Stability of Water Drives in Water-Wet Connate-Water-Bearing Reservoirs},
    journal = {Society of Petroleum Engineers Journal},
    volume = {14},
    number = {01},
    pages = {63-74},
    year = {1974},
    month = {02},
    issn = {0197-7520},
    doi = {10.2118/4268-PA},
}

@article{yortsos1989,
author = {Yortsos, Y. C. and Hickernell, F. J.},
title = {Linear Stability of Immiscible Displacement in Porous Media},
journal = {SIAM Journal on Applied Mathematics},
volume = {49},
number = {3},
pages = {730-748},
year = {1989},
doi = {10.1137/0149043},
URL = {https://doi.org/10.1137/0149043}
}

@article{chikhliwala1988,
author = {Chikhliwala, E. and Huang, A. and Yortsos, Yannis},
year = {1988},
month = {06},
pages = {257-276},
title = {Numerical study of the linear stability of immiscible displacement in porous media},
volume = {3},
journal = {Transport in Porous Media},
doi = {10.1007/BF00235331}
}

@article{riaz2003,
author = {Riaz, Amir and Meiburg, Eckart},
year = {2003},
month = {04},
pages = {},
title = {Radial source flows in porous media: Linear stability analysis of axial and helical perturbations in miscible displacements},
volume = {15},
journal = {Physics of Fluids - PHYS FLUIDS},
doi = {10.1063/1.1556292}
}

@article{pinilla2021,
title = {Experimental and computational advances on the study of Viscous Fingering: An umbrella review},
journal = {Heliyon},
volume = {7},
number = {7},
pages = {e07614},
year = {2021},
issn = {2405-8440},
doi = {https://doi.org/10.1016/j.heliyon.2021.e07614},
author = {Andrés Pinilla and Miguel Asuaje and Nicolás Ratkovich}
}

@article{manickam1995, 
title={Fingering instabilities in vertical miscible displacement flows in porous media},
volume={288},
DOI={10.1017/S0022112095001078},
journal={Journal of Fluid Mechanics},
author={Manickam, O. and Homsy, G. M.},
year={1995},
pages={75-102}
}

@article{dewit1997,
    author = {De Wit, A. and Homsy, G. M.},
    title = {Viscous fingering in periodically heterogeneous porous media. I. Formulation and linear instability},
    journal = {The Journal of Chemical Physics},
    volume = {107},
    number = {22},
    pages = {9609-9618},
    year = {1997},
    month = {12},
    issn = {0021-9606},
    doi = {10.1063/1.475258},
    url = {https://doi.org/10.1063/1.475258},
}

@article{scovazzi12017,
title = {Analytical and variational numerical methods for unstable miscible displacement flows in porous media},
journal = {Journal of Computational Physics},
volume = {335},
pages = {444-496},
year = {2017},
issn = {0021-9991},
doi = {https://doi.org/10.1016/j.jcp.2017.01.021},
url = {https://www.sciencedirect.com/science/article/pii/S0021999117300311},
author = {Guglielmo Scovazzi and Mary F. Wheeler and Andro Mikelić and Sanghyun Lee}
}

@book{freeze1979,
author = {Freeze, R. Allan and Cherry, John A.},
title = {Groundwater}, publisher = {Prentice-Hall},
year = {1979},
address = {Englewood Cliffs, N.J.},
isbn = {978-0133888300},
}

@book{pedersen2006,
author = {Pedersen, Karen Schou and Christensen, Peter Lindskou and Shaikh, Jawad Azeem and Christensen, Peter L.},
title = {Phase Behavior of Petroleum Reservoir Fluids},
publisher = {CRC Press},
year = {2006},
address = {Boca Raton},
isbn = {1420018256},
doi = {https://doi.org/10.1201/9781420018257},
}

@article{riaz2006, 
author={Riaz, A. and Hesse, M. and Tchelepi, H. A. and Orr, F. M.}, 
title={Onset of convection in a gravitationally unstable diffusive boundary layer in porous media}, 
volume={548}, 
journal={Journal of Fluid Mechanics}, 
year={2006}, 
pages={87–111},
doi={10.1017/S0022112005007494}, 
}

@article{lasser2021,
title={Stability and dynamics of convection in dry salt lakes},
volume={917},
DOI={10.1017/jfm.2021.225},
journal={Journal of Fluid Mechanics},
author={Lasser, Jana and Ernst, Marcel and Goehring, Lucas},
year={2021},
pages={A14},
}

@article{vanduijn2019,
author    = {van Duijn, C. J. and Pieters, G. J. M. and Raats, P. A. C.},
title     = {On the Stability of Density Stratified Flow Below a Ponded Surface},
journal   = {Transport in Porous Media},
year      = {2019},
volume    = {127},
number    = {3},
pages     = {507--548},
month     = {04},
doi       = {10.1007/s11242-018-1209-9},
url       = {https://doi.org/10.1007/s11242-018-1209-9},
issn      = {1573-1634}
}

@article{bringedal2025,
author    = {Bringedal, C. and Kiemle, S. and van Duijn, C. J. and Helmig, R.},
title     = {Impact of Saturation on Evaporation-Driven Density Instabilities in Porous Media: Mathematical and Numerical Analysis},
journal   = {Transport in Porous Media},
year      = {2025},
volume    = {152},
number    = {10},
pages     = {72},
month     = {08},
doi       = {10.1007/s11242-025-02207-y},
url       = {https://doi.org/10.1007/s11242-025-02207-y},
issn      = {1573-1634}
}

\end{document}